\documentclass[preprint,aps,prd,nofootinbib,floatfix,amsmath,amssymb]{revtex4-2}

\usepackage{graphicx}%
\usepackage{multirow}%
\usepackage{pdfsync}
\usepackage{amsthm}%
\usepackage{mathrsfs}%
\usepackage{xcolor}%
\usepackage{float}
\usepackage{subfig}
\usepackage{easyReview}
\usepackage{hyperref}

\begin{document}
	
\title[The decay of the massive boson]{Decay of the $Z^\prime$  gauge boson with lepton flavor violation}
\author{D. Espinosa-G\'{o}mez$^{1,2}$\footnote{despinosa3187@gmail.com},
		F. Ram\'{i}rez-Zavaleta$^1$\footnote{iguazu.ramirez@umich.mx},  E. S. Tututi$^1$\footnote{eduardo.tututi@umich.mx}}
\address{$^1$Facultad de Ciencias F\'{i}sico Matem\'{a}ticas,
		Universidad Michoacana de San Nicol\'{a}s de Hidalgo, Avenida Francisco J. M\'{u}jica S/N, 58060, Morelia, Michoac\'{a}n, M\'{e}xico}
\address{$^2$Ingenier\'ia en Tecnolog\'ias de la Informaci\'on y Comunicaciones, Instituto Tecnol\'ogico Superior de Puru\'andiro, Carretera Puru\'andiro-Galeana km 4.3, 58532, Mpio. de Puru\'andiro, Michoac\'an, M\'exico}

\begin{abstract}
The flavor-violating decay of a new neutral massive gauge boson $Z^\prime\to\mu e$   is analyzed in the context of  extended models, in which this particle emerges. By means of the analysis of the $\mu\to e\gamma$ decay,  $\mu-e$ conversion process in nuclei and the $\mu \to e e^{+}e^{-}$ decay, the strength of the $Z^\prime\mu e$ coupling is estimated and  used to calculate   the branching ratio of the $Z^\prime\to \mu e$  decay.
This is done for the so-called $Z_S,\,Z_{LR},\,Z_\chi, Z_\psi$ and $Z_\eta$ bosons. We found that,  through the $\mu-e$ conversion process,  the most restrictive bound for the coupling is provided by the $Z_{LR}$ boson. Meanwhile, by means of the $\mu \to e e^{+}e^{-}$ decay,  the most restrictive  bound  for the $Z^\prime\mu e$ coupling is provided by the $Z_\chi$ boson. However, if we concentrate on  the  less restrictive prediction for the Br($Z^\prime\to \mu e$),  this comes from the $Z_\eta$ boson and the resulting branching ratio is less than $10^{-4}$. On the other hand, if we consider the most restrictive bound,  the branching ratio for the process is below $2\times 10^{-7}$, which results from the   $Z_{RL}$ boson, and is obtained through the $\mu-e$ conversion process.
\end{abstract}


\maketitle

\section{Introduction}\label{sec1}

It is well known that   Flavor  Changing Neutral Currents (FCNC) produce  very suppressed flavor transition processes in the quark sector of the  Standard Model (SM). Additionally to a factor of suppression due to the GIM mechanism, these  transitions are induced  at  one-loop level~\cite{SMqsectorsup}. Beyond the standard model (SM), there are many types  of new physics (NP) that can lead FCNC, even at the tree level. These types of  NP become important since  associated transitions could be enhanced at  tree level  and consequently competitive with respect to the  corresponding ones in  the SM at the one-loop level.  Moreover, within the  NP context,  it could arise effects that are definitively forbidden in the SM ~\cite{becker,apollonio,fukuda}.

It draws the attention that even with minimal extensions to the SM, for example, by including right-handed components of neutrinos fields  into it, some non-trivial physical properties can be explained, such as the neutrino oscillations, which preserve lepton number  only at the global level~\cite{becker,apollonio,fukuda}. Nevertheless, within the frame of  this extension, some predictions result quite suppressed and  the theory requires the use of  more elaborated extensions that could  include lepton flavor violation  in its Lagrangian.
Concretely, the  branching ratio  Br$(\tau\to\mu\gamma)<10^{-40}$, that is calculated with minimal extensions to the SM~\cite{cheng-li}, is very suppressed with respect to the experimental bound Br$(\tau\to\mu\gamma)<10^{-8}$ reported in the Particle Data Group (PDG)~\cite{pdg2022}.

An interesting   process that violates the flavor symmetry in the lepton sector that has been  studied in different scenarios is the  $Z\to \bar{l}_il_j$ ($l_i\neq l_j=e,\mu,\tau$) decay~\cite{mendez-mir-otros}. Although some   branching ratios of these decays are more suppressed with respect to the corresponding  experimental  bounds, others are of the same order of magnitude \cite{pdg2022}.
Similar flavor-violating decays involving a new neutral  massive gauge boson, denoted as $Z^\prime$ \cite{robinet1,lang2,Zwin,JRizo, perez-soriano,framton,leike,arhrib,lan2009,aabo,siru},  show that the branching ratio of these processes   could be less suppressed than the respective decays  of the  $Z$ boson.   Indeed, in Ref. \cite{aranda2012}  the  flavor-violating decay   $Z^\prime\to\tau\mu$  has been studied in models where the so-called $Z_S, Z_{LR}, Z_\chi, Z_\psi$ and $Z_\eta$ bosons arise. It was found that the most restrictive branching ratio for the process in question is less than $10^{-2}$.

On the other hand, the CMS and ATLAS~\cite{FVcms,FVatlas} collaborations have been  searching for new heavy particles via lepton flavor violation, in particular, the $Z^\prime$ boson decaying into pair of leptons $\mu e, \tau e, \tau\mu$. Especially, the CMS~\cite{FVcms} collaboration  performed an analysis on the results of the cross-section and the branching ratios of the $Z^\prime$ boson decaying into $\mu e$ in proton-proton collisions at the CM energy $\sqrt{s}=13$ TeV with an integrated luminosity of 35.9 fb$^{-1}$; no evidence on physics beyond the SM related to the $\mu e$ spectrum was found. The ATLAS~\cite{FVatlas} collaboration  studied the  cross-section times the branching ratio of the $Z^\prime\to \mu e, \tau e, \tau\mu$ decays as a function of the $Z^\prime$ boson mass, in proton-proton collisions with an integrated luminosity of 36.1 fb$^{-1}$.  As before, none evidence on excess over the SM predictions was found. The CMS Collaboration concludes that a  $Z^\prime$ boson with a 10\%  branching ratio to the $\mu e$ channel is excluded for masses below 4.4 TeV \cite{FVcms}. While, the ATLAS Collaboration concludes that, from the $e \mu$, $e\tau$ and $\mu\tau$ final states, Bayesian lower limits at 95 \% credibility level on the mass
of a $Z^\prime$ gauge boson with lepton flavor-violating (LFV) couplings are set at 4.5, 3.7, and 3.5 TeV, respectively~\cite{FVatlas}. However, to completely discard these lepton flavor-violating processes, it is necessary to increase the confidence level on the measurements  of such decays as well as the mass search range of the $Z^\prime$ gauge boson. Hence, there is still an open window for theoretical analysis. In fact, a calculation of these flavor-violating decays, in a broad mass interval could be helpful  to exclude or  leave open the possibility of the existence of the $Z^\prime$ gauge boson.

In this work, we study the lepton flavor violation  through the   $Z^\prime \to\mu e$  decay process, by using several models that predict a $Z^\prime$ gauge boson. This is done by using the Package-X software~\cite{PX} that facilitates the calculation  of the Feynman diagrams involved in the process, since  it is possible to obtain analytical expressions for the amplitudes under analysis.
The process in question  is  analyzed and bounded by using the $\mu\to e\gamma$ decay, the $\mu-e$ conversion process in nuclei and the $\mu \to e e^{+}e^{-}$ decay~\cite{pdg2022}, which are useful  for extracting bounds of the  $Z^\prime\mu e$ coupling and then to constrain the decay under study.
In fact, our approach  introduces the $\mu e$ decay mode promoted by the presence of the $Z^\prime$ boson as a feasible possibility to be considered in direct searches. However, complementarily, an experimental spin analysis would be required in order to establish or discard it.

The organization of this paper is as follows: Section~\ref{framea} describes the flavor-violating Lagrangian in the context of extended models. In section~\ref{decaymugamma}, we analyze the $\mu\to e\gamma$ decay mediated by a $Z^\prime$ boson along with the $\mu-e$ conversion process in nuclei and the $\mu \to e e^{+}e^{-}$ decay to  estimate the strength of the $Z^\prime\mu e$ coupling. In section~\ref{decays}, we present a discussion on the $Z^\prime\to \mu e$ decay.  Finally, in section~\ref{conclu} the conclusions are presented.

\section{Theoretical framework}\label{framea}

\subsection{The different models}

There are several  extended models that predict  the presence of new massive neutral gauge bosons, commonly denoted as $Z^\prime$. Here,  we  describe briefly the models  used in this work. The  grand unified group $E_6$~\cite{robinet1,Zwin,lang2}  breaks directly: $E_6\to SO(10)\times U(1)_\psi$ and  $SO(10)\to SU(5)\times U(1)_\chi$. In general, the $E_6$ gauge symmetry group predicts at most four neutral gauge bosons, however,  we only consider the  $Z_\psi$ boson and the $Z_\chi$ gauge boson that arises from symmetry breaking  of the group $SO(10)$.
On the other hand, Hewett and Rizzo~\cite{JRizo} established that the group $U(1)_\psi\times U(1)_\chi$ can be reduced to an effective group $U(1)_\theta $  ($U(1)_\psi\times U(1)_\chi\to U(1)_\theta$), where $U(1)_\theta$ is a linear combination of $U(1) _\psi$ and $U(1)_\chi$. The gauge fields $Z_\psi$ and $Z_\chi$ corresponding to $U(1)_\psi$ and $U(1)_\chi$ symmetries respectively, are massive and the mass eigenstates are defined as~\cite{JRizo,Gi98,PJW}
\begin{equation}
	Z^\prime(\theta)\equiv Z_\chi \text{cos}(\theta)+Z_\psi \text{sin}(\theta).
\end{equation}
By varying $\theta$, it can be identified  different bosons  $Z^\prime$, for instance: the $Z_\chi$ boson with $\theta=0$, the $Z_\psi$ boson with $\theta=\pi/2$. In particular,  with $\theta=\text{tan}^{-1}(-\sqrt{5/3})$, the  $Z_\eta$ boson results as a linear combination of the $Z_\chi$ and  $Z_\psi$ gauge bosons, given explicitly  $Z_\eta=\sqrt{\frac{3}{8}}Z_\chi-\sqrt{\frac{5}{8}}Z_\psi$. These three bosons arise in many superstring-inspired models in which the $E_6$ group breaks directly to a rank-5 group \cite{Zwin,lang2}.  At the last stage of  the breaking symmetry of grand unification theories, other that $SU(5)$,  the breaking pattern generally  finishes   with at least one neutral gauge boson $Z^\prime$,
it is  expected that its mass  be of order of  TeV's. This situation can be described by the theory  based on the  symmetry group $SU(2)_L\times U(1)\times U^{\prime}(1)$. Following Refs.~\cite{durkin,arhrib},  in the gauge eigenstate basis, the neutral current Lagrangian   can be written as
\begin{equation}
{\cal L}_{NC}=-eJ_{EM}^\mu A_{\mu}-g_1J^{\mu(1)}Z_{1\mu}-g_2J^{\mu(2)}Z_{2\mu},
\label{ncg}
\end{equation}
where $J_{EM}^\mu$ is the electromagnetic current,  $Z_1$ is the $SU(2)\times U(1)$ neutral gauge boson, and $Z_2$ is the new gauge boson associated with the additional Abelian symmetry  $U^{\prime}(1)$. The $g_{1,2}$  couplings  are of the form
\begin{equation}
	g_2= \sqrt{\frac{5}{3}} \sin \theta_W g_1 \lambda_g, \label{1b}
\end{equation}
where $g_1=g/\cos \theta_W$,  being $g$  the weak coupling,  and $\theta_W$  the Weinberg angle and $\lambda_g$ depends of the symmetry breaking pattern, which is usually assumed $\mathcal{O}(1)$~\cite{robinet1}.
For simplicity, it  can be assumed that there is no mixing between $Z_1$ and $Z_2$, and consequently they are the mass eigenstates of $Z$ and $Z^\prime$, respectively. The current associated with the group $U^{\prime}(1)$  can be cast as
\begin{equation}
J^{\mu(2)}=\sum_{i}\overline{f}_i\gamma^{\mu}\left(\epsilon_{f_{i L}}P_L+\epsilon_{f_{i R}}P_R\right)f_i,
\label{current2}
\end{equation}
where  $i$ runs over all quarks and leptons and $P_{L,R}=\frac{1}{2}(1\mp\gamma^5)$ are the quiral projectors. The quiral couplings are $\epsilon_{f_{iL,R}}=Q^{f_i}_{L,R}$  where the various $Q's$ are the quiral charges. For the case of the $Z_\psi$, $Z_\chi$ and $Z_\eta$ bosons, the chiral charges were determined in \cite{leike}.

The $Z_{LR}$ gauge boson emerges in the left-right symmetric model in a scheme of breaking symmetry just as $SO(10)\to SU(3)\times SU(2)_L\times SU(2)_{R}\times U(1)_{B-L}$~\cite{durkin,RW82,mc,langacker3,Fa}.
Let us consider the  model in which $Z_{LR}$ is orthogonal to the $Z_1$ in Eq. (\ref{ncg}) that couples to the
 "left-right" current
\begin{equation}\label{cLR}
	J_{LR}^{\mu}\equiv \sqrt{\frac{3}{5}}\big(\alpha J_{3R}^{\mu}-(1/2\alpha)J_{B-L}^{\mu}\big),
\end{equation}
where the currents $J_{3R}^{\mu}$ and  $J_{B-L}^{\mu}$ are associated with the third component of $SU(2)_R$ and $B-L$ symmetry, respectively; $B(L)$ denotes the barionic (leptonic) number and $\alpha=\sqrt{(1-2\sin^2\theta_W)/\sin^2\theta_W}\approx 1.53$. The $J_{3R}^{\mu}$ current is constructed  in such a way that all the right-handed fermions are doublets and all the left-handed fermions are singlets.
In this way,  the chiral couplings are  given  by~\cite{leike,lang2}:
\begin{align}
\epsilon^{LR}_{{f_i}L}=&\sqrt{3/5}\left(-\frac{1}{2\alpha}\right)(B-L)_{f_i},\nonumber\\
\epsilon^{LR}_{{f_i}R}=&\sqrt{3/5}\left(\alpha T_{3R}^{f_i}-\frac{1}{2\alpha}\right)(B-L)_{f_i},
\end{align}
where $T_{3R}^{f_i}$ is the third component of its right-handed isospin in the $SU(2)_R$ group.

The sequential $Z_S$ boson is defined to have the same couplings to fermions as the SM  $Z$ boson. In this model, $g_2 =g_1$. It is useful as reference when comparing constraints from  different sources.

Finally, the chiral charges $Q^{fi}_{L,R}$ of the different models aforementioned are obtained and discussed in Refs.~\cite{leike,lang2,arhrib,RW82,Fa}; these  are summarized in Table~\ref{tablaquiral}.

\begin{table}[htbp]
\caption{Chiral-diagonal couplings of the extended models.}
\begin{center}
\begin{tabular}{|r|r|r|r|r|r|r|r|}
\toprule
& $Q^{u}_L$ & $Q^{u}_R$ & $Q^{d}_L$ & $Q^{d}_R$ &$Q^{e}_L$ & $Q^{e}_R$ &$Q^{\nu}_L$\\ \hline
$Z_S$    &  0.3456   & -0.1544   & -0.4228   & 0.0772    & -0.2684  &0.2316     &0.5 \\ \hline			
$Z_{LR}$ &-0.08493   & 0.5038    &-0.08493   &-0.6736    & 0.2548   &-0.3339    &0.2548 \\ \hline			
$Z_{\chi}$ &$\frac{-1}{2\sqrt{10}}$   &$\frac{1}{2\sqrt{10}}$    &$\frac{-1}{2\sqrt{10}}$   &$\frac{1}{2\sqrt{10}}$    &$\frac{3}{2\sqrt{10}}$   &$\frac{-3}{2\sqrt{10}}$    &$\frac{3}{2\sqrt{10}}$ \\ \hline			
$Z_{\psi}$ &$\frac{1}{\sqrt{24}}$   &$\frac{-1}{\sqrt{24}}$    &$\frac{1}{\sqrt{24}}$   &$\frac{-1}{\sqrt{24}}$    &$\frac{1}{\sqrt{24}}$   &$\frac{-1}{\sqrt{24}}$    &$\frac{1}{\sqrt{24}}$ \\ \hline	
$Z_{\eta}$ &$\frac{-2}{2\sqrt{15}}$   &$\frac{2}{2\sqrt{15}}$    &$\frac{-2}{2\sqrt{15}}$   &$\frac{2}{2\sqrt{15}}$    &$\frac{1}{2\sqrt{15}}$   &$\frac{-1}{2\sqrt{15}}$    &$\frac{1}{2\sqrt{15}}$ \\ 
\botrule
\end{tabular}
\label{tablaquiral}
\end{center}
\end{table}

\subsection{The generic Lagrangian}

In  order to study  the $Z^\prime\to \mu e$  decay,  we  use of the most general renormalizable Lagrangian that includes flavor violation mediated by the $Z^\prime$ gauge boson. Such a Lagrangian can be written as follows \cite{durkin,langacker3,Salam-Mohapatra}:
\begin{equation}
	\mathcal{{L}}_{NC}=\sum\limits_{ij}\Bigg[\bar{f}_{i} \gamma^\alpha(\Omega_{Lfifj}\,P_L+\Omega_{Rfifj}\,P_R)f_j+ \bar{f}_{j}\gamma^\alpha (\Omega^{*}_{Lfifj}\,P_L+
	\Omega^{*}_{Rfifj}\,P_R)f_i\Bigg] Z^{\prime}_{\alpha}, \label{1a}
\end{equation}
where $f_{i}$ is any fermion of the SM and $Z^{\prime}_{\alpha}$ is the neutral massive gauge boson predicted by several extensions of the SM~\cite{robinet1,lang2,perez-soriano,framton,leike,arhrib,lan2009}. Thus, the $\Omega_{Ll_i l_j}$ and $\Omega_{Rl_i l_j}$ parameters represent the strength of the $Z^\prime l_i l_j$ coupling, where $l_{i}$ is any lepton of the SM. We will assume that the couplings are symmetric: $\Omega_{R, L\,l_i l_j}=\Omega_{R, L\,l_j l_i}$. It is convenient to express the diagonal components of the $\Omega$ parameter in terms of the quiral charges as follows: $\Omega_{Lfifi}= -g_2 Q^{fi}_L$ and $\Omega_{Rfifi}= -g_2 Q^{fi}_R$, where the various $Q^{fi}_{L,R}$ are the quiral charges,  shown in Table~\ref{tablaquiral}.
The non-diagonal the $\Omega_{Ll_i l_j}$ and $\Omega_{Rl_i l_j}$ parameters  are to be somehow determined.

\section{The processes}\label{decaymugamma}
\subsection{The $\mu\to e\gamma$ decay}

\begin{figure}[htb!]
	\begin{center}
		\includegraphics[scale=0.5]{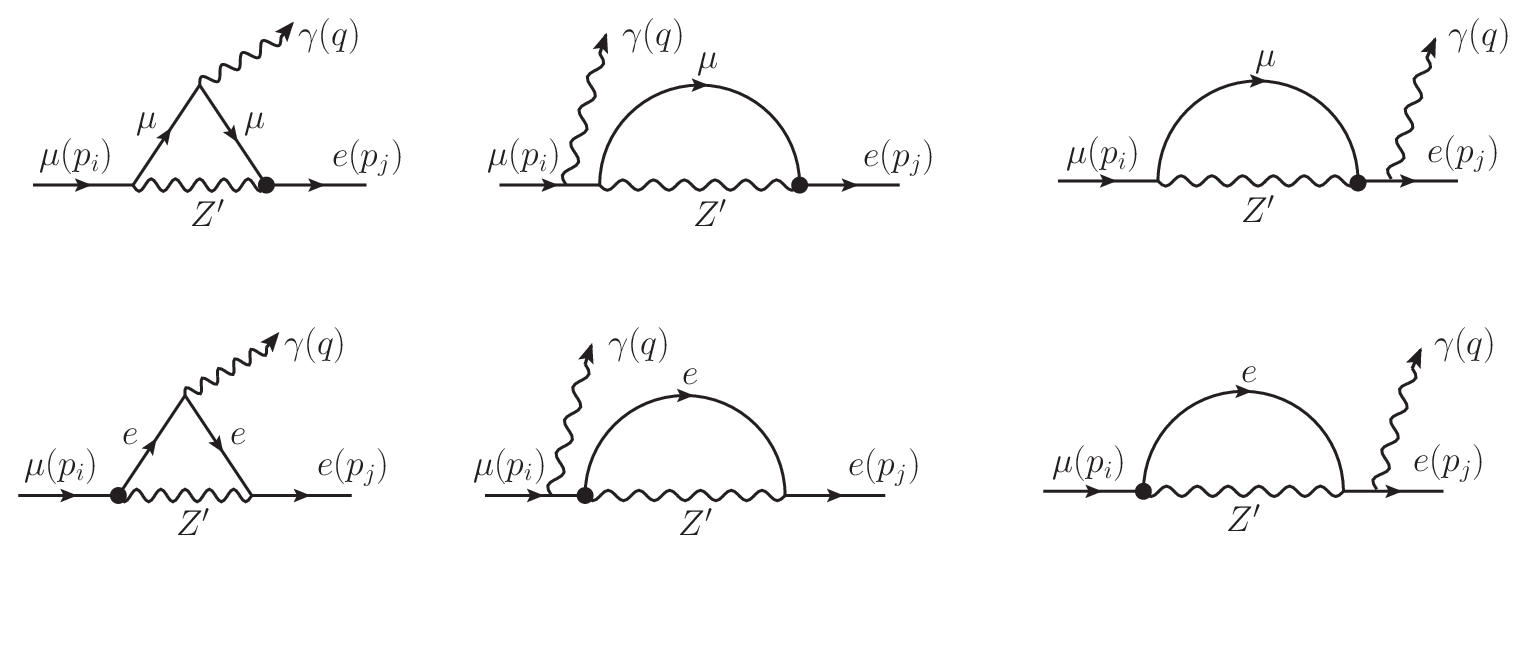}
		\caption{Feynman diagrams contributing to the  $\mu\to e\gamma$ decay.}\label{de3}
	\end{center}
\end{figure}

The   Feynman diagrams corresponding to this decay  are shown in Fig.~\ref{de3}, where  the fermionic  internal line represents a muon or an electron.  In order to keep dominant contributions, we only consider  diagrams  involving just one vertex that induces lepton  flavor violation.
Specifically, the associated Feynman rules with the $Z^\prime e e$ and $Z^\prime\mu e$ vertices are $i\gamma^{\alpha}g_2(Q_{L}^e P_L+Q_{R}^e P_R)$ and $-i\gamma^{\alpha}(\Omega_{L\mu e} P_L+\Omega_{R\mu e} P_R)$, respectively.  This vertex rule is denoted in the Feynman diagram with a dot.
If we consider two vertices inducing lepton flavor violation, then  it is necessary to consider  loop diagrams with  internal fermionic lines  that corresponds to a tau lepton. Thus, the amplitudes corresponding to these diagrams are proportional to $\Omega_{\mu \tau}\Omega_{\tau e}$ or to $\Omega_{\mu e}^2$, since $\Omega_{\mu \tau}\Omega_{\tau e}\approx \Omega_{\mu e}^2$ \cite{david}.   The  resulting amplitudes are suppressed by a factor $\Omega_{\mu e}$ ($\sim 10^{-3}-10^{-2}$  in the interval $m_{Z^\prime}=[2,7]$ TeV) with respect to the  amplitudes obtained by using only one lepton  flavor violation vertex and they can be neglected.
Notice that  the calculation of the amplitude is carried out in the context of the unitary gauge. Thus, we find that the total amplitude can be expressed as
\begin{equation}
	\mathcal{M}(\mu\to e\gamma)=\bar{u}(p_j) \sigma^{\mu\alpha} q_{\alpha} \:\big(F_M+F_E \gamma^5\big)\,u(p_i)\epsilon^{*}_{\mu}(q), \label{amplifoton}
\end{equation}
where
\begin{equation}
	F_{M}=\frac{ieg_2}{64 \pi^2 m_{\mu}} \Bigg[F_1(Q_L^e-Q_R^e)\big(\Omega_{L\mu e}-\Omega_{R\mu e})+F_2(Q_L^e\Omega_{L\mu e}+Q_R^e\Omega_{R\mu e}\big)\bigg], \label{FFfoton1}
\end{equation}
\begin{equation}
	F_{E}=\frac{ieg_2}{64 \pi^2 m_{\mu}} \Bigg[F_1(Q_L^e-Q_R^e)\big(\Omega_{L\mu e}+\Omega_{R\mu e})+F_2(Q_L^e\Omega_{L\mu e}-Q_R^e\Omega_{R\mu e}\big)\bigg], \label{FFfoton2}
\end{equation}
being $m_\mu$ the   muon lepton mass. As usual, $e$ represents the electric charge of the electron. The form factors $F_1$ and $F_2$ are explicitly given in the Appendix. It should be noted that the structure of these form factors was obtained using Package-X~\cite{PX}. The amplitude  in Eq.~(\ref{amplifoton}) is free of ultraviolet divergences.
According to the definition of  decay width~\cite{pdg2022},   we obtain the branching ratio for the process in question
\begin{eqnarray}
	\textup{Br}(\mu\to e\gamma) &=& \frac{\alpha g_2^2}{4096 \pi^4}\bigg[|F_1 (Q_L^e-Q_R^e)+F_2Q_L^e|^2 |\Omega_{L\mu e}|^2\nonumber \\
	&&+|F_1(Q_R^e-Q_L^e)+F_2 Q_R^e|^2 |\Omega_{R\mu e}|^2 \bigg] \frac{m_{\mu}}{\Gamma_{\mu}}, \label{bramu}
\end{eqnarray}
where  $\Gamma_{\mu}$ is the total decay width of the muon lepton.

\subsection{The $\mu-e$ conversion process}
The  $\mu-e$ conversion process refers to the capture of a muon  by a heavy nucleus, in which the lepton flavor conservation is not preserved~\cite{kuno,af,kosmas}.  Such processes is represented by
\begin{equation}
	\mu^{-}+N \to e^{-}+N,
\end{equation}
where the nucleus $N$ has  mass and atomic number $A$ and $Z$, respectively.
\begin{figure} [h!]
	\centering
	\includegraphics[scale=.7]{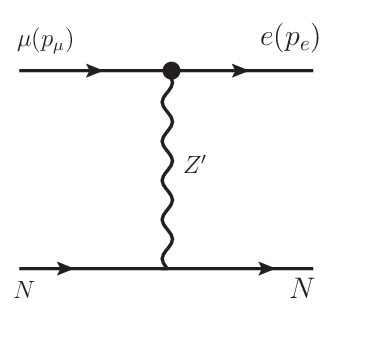}
	\caption{ Feynman diagram contributing to the $\mu-e$ conversion process, non-photonic case.}\label{DFconversion}
\end{figure}
In order to obtain a useful branching ratio for this process, we focus on the case of neutral current interactions with flavor violation
at the tree level mediated by a $Z^\prime$ gauge boson (see Fig.~\ref{DFconversion}). Specifically, we
analyze the non-photonic process described the lepton-quark effective Lagrangian~\cite{Jbernabeu,kosmas}
\begin{equation}
\mathcal{L}_{eff}=\sqrt{2}\,G\,\bar{u}(p_e) \gamma^{\lambda}\big(g_V^\prime-g_A^\prime \gamma_5\big) u(p_\mu) \sum_{q=u,d,s..}\bar{q}\gamma_\lambda\bigg(v_q^\prime-a_q^\prime \gamma_5\bigg)q, \label{eff}
\end{equation}
with the various parameters defined as
\begin{align} \label{eacopling}
G&=\frac{g_2}{\sqrt{32}m_{Z^\prime}^2},\nonumber\\
g_V^\prime&=\Omega_{L\mu e}+\Omega_{R\mu e},\, g_A^\prime=\Omega_{R\mu e}-\Omega_{L\mu e},\nonumber\\
v_q^\prime&=Q_{L}^{q_i}+Q_{R}^{q_i},\,  a_q^\prime=Q_{R}^{q_i}-Q_{L}^{q_i}.
\end{align}
After  algebraic manipulations and introducing the nucleon spinor $\psi_N=(^p_n)$ and the Pauli matrix $\tau_3$, Eq.~(\ref{eff}) can be rewritten as
\begin{align}
	\mathcal{L}_{eff}^N=&\sqrt{2}\,G\,\bar{u}(p_e) \gamma^{\lambda}\big(g_V^\prime-g_A^\prime \gamma_5\big) u(p_\mu)\nonumber \\
	& \times \bar{\psi}_N\gamma_\lambda\big[(C_{1S}+C_{1V} \tau_3)-(C_{2S}+C_{2V}\tau_3)\gamma_5\big]\psi_N, \label{effN}
\end{align}
where the couplings $C_{1S}$ and $C_{1V}$ are denoted as the isoscalar vector and the isovector vector, respectively. These parameters are given explicitly as
\begin{align}
	C_{1S}\equiv&\frac{3}{2}\big(v_u^\prime+v_d^\prime\big),\\
	C_{1V}\equiv&\frac{1}{2}\big(v_u^\prime-v_d^\prime\big).
\end{align}
On the other hand, in Refs.~\cite{kosmas,HC2} it has been shown that the couplings of axial quarks $C_{2S}$ and $C_{2V}$ do not contribute to the coherent nuclear charge ($Q_{W }^{\prime}$), therefore the width of the $\mu-e$conversion process  does not depend on these parameters. Thus, the width can be written as
\begin{align}
	\Gamma=\frac{g_2^2 \alpha^3 m_\mu^5}{32\pi^2 m_{Z^\prime}^4}\frac{Z_{eff}^4}{Z}|F(q)|^2 Q_{W}^{\prime\,2} \big(g_{V}^{\prime\,2}+g_{A}^{\prime\,2}\big).\label{widthmu}
\end{align}
The most stringent bounds for capture rate   are provided by gold and titanium nuclei \cite{pdg2022}.
The branching ratio for the $\mu -e$~\cite{kuno} conversion is given by
\begin{equation}
	\mathrm{Br}(\mu^{-}+N\to e^{-}+N) \equiv \frac{\Gamma(\mu^{-}+N \to e^{-}+N)}{\Gamma_{caputure}},\label{1e}
\end{equation}
where $\Gamma_{caputure}\equiv\Gamma(\mu^{-}+N \to capture)$ is the muon  decay width. For our propose of bounding the $\Omega_{R,L{\mu e}}$ couplings, we use the branching ratio for the $\mu-e$ conversion, which, according to Eq. (\ref{widthmu}), can be expressed as
\begin{equation}
	\mathrm{Br}(\mu^-+ N\to e^- +N)=\frac{g_2^2\alpha^3 m_\mu^5}{16 \pi^2 m_{Z^\prime}^4}\frac{Z_{eff}^4}{Z}|F(q)|^2\big(|\Omega_{L\mu e}|^2+|\Omega_{R\mu e}|^2\big)\frac{Q_{W}^{\prime 2 }}{\Gamma_{capture}}, \label{branchingMue}
\end{equation}
with
\begin{equation}
	Q_{W}^{\prime 2}=(2Z+N)(Q_L^u+Q_R^u)+(Z+2N)(Q_L^d+Q_R^d). \label{qw}
\end{equation}
We consider the $Z^\prime$ gauge boson for the different models aforementioned.
However,  as it can be appreciated from  Eqs. (\ref{branchingMue}) and (\ref{qw}),  the resulting branching ratio is proportional to the quiral charges of the quarks $u$ and $d$, which  vanishes for the $Z_{\chi}, Z_\psi$ and $Z_\eta$ bosons, since  $Q_L^q+Q_R^q=0$ \cite{robinet1,brenda}. For the remaining bosons, $Z_S$ and $Z_{LR}$, we take the following parameters of the titanium nucleus: $Z_{eff}\simeq 17.6$, $F(q^2\simeq-m_{\mu}^2)\simeq 0.54$ and $\Gamma_{capture}=(2.59\pm 0.012)\times 10^{-6} s^{-1}$~\cite{Jbernabeu, Suzuki, HC2, HC}, while the experimental bound for $\mathrm{Br}(\mu^{-}+N\to e^{-}+N)_{\mathrm{exp}}<4.3\times 10^{-12}$~\cite{sundrumII}.
Eqs. (\ref{bramu}) and (\ref{branchingMue}) will be our starting point in next subsection  to constrain the  $\Omega_{R,L \mu e}$ couplings.

\subsection{The $\mu \to e e^{+}e^{-}$ decay}

\begin{figure}
	\centering
	\includegraphics[scale=0.7]{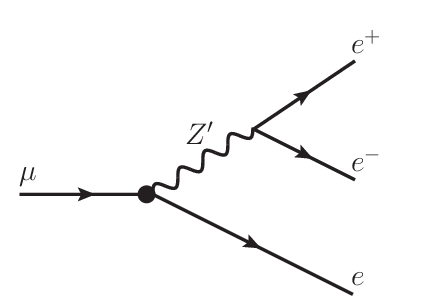}
	\caption{Feynman diagram for the $\mu\to ee^+e^-$ decay process.}\label{three-body-decay}
\end{figure}

The $\mu \to e e^{+}e^{-}$ process can be directly calculated at the tree level.
The corresponding Feynman diagram  is shown in Fig. \ref{three-body-decay}.
By following Ref.~\cite{aranda2012}, we found that Br($\mu \to e e^{+}e^{-}$)  can be written as
\begin{align}
\mathrm{Br}(\mu \to e e^{+}e^{-})  =&\frac{g_2^2}{384\pi^3} \left[f(\frac{m_{Z^\prime}^2}{m_\mu^2})\left(|Q_L^e\Omega_{L\mu e}|^2+|Q_R^e\Omega_{R\mu e}|^2\right)\right.\nonumber\\
&\left. + h(\frac{m_{Z^\prime}^2}{m_\mu^2})\left(|Q_L^e\Omega_{R\mu e}|^2+|Q_R^e\Omega_{L\mu e}|^2\right) \right]\frac{m_\mu}{\Gamma_\mu}
\label{three-decay},
\end{align}
where
\begin{align*}
f(a)=&\int_0^1 dx \frac{2x-1}{(x-1+a)^2}(2(7-4x)x-5)\approx \frac{1}{a},\\
 h(a)=& \int_0^1 dx \frac{2x-1}{(x-1+a)^2}(1-2(x-1)x)\approx \frac{1}{a^2}.
\end{align*}
In last equations, our  approximations are valid for $a = \frac{m_{Z^\prime}^2}{m_\mu^2}>>1$, which stand for $m_{Z^\prime}>$ 1 TeV.  Thus, the function $h(a)$ is  suppressed with respect to the function $f(a)$ by at lest 8 orders of magnitude. So, we can can drop the  term, $h(\frac{m_{Z^\prime}^2}{m_\mu^2})\left(|Q_L^e\Omega_{R\mu e}|^2+|Q_R^e\Omega_{L\mu e}|^2\right)$ in (\ref{three-decay}) and this simplifies the bounding process of the couplings. In this manner the final expression for the branching ratio is expressed as
\begin{align}
\mathrm{Br}(\mu \to e e^{+}e^{-}) \approx \frac{g_2^2}{384\pi^3} f(\frac{m_{Z^\prime}^2}{m_\mu^2})\left(|Q_L^e\Omega_{L\mu e}|^2+|Q_R^e\Omega_{R\mu e}|^2\right)\frac{m_\mu}{\Gamma_\mu}.
\label{three-decay-bis}
\end{align}

\subsection{Bounding the $\Omega_{\mu e}$ coupling }

As it was mentioned, we are interested in bounding the flavor-violating  branching ratio of the $Z^\prime$ gauge boson decaying into the $\mu$ and $e$ leptons.  In order to do that,  we firstly bound the $|\Omega_{R,L \mu e}|^2$ couplings by using the branching ratios in Eqs. (\ref{bramu}),  (\ref{branchingMue}) and (\ref{three-decay-bis}). The task is carried out by considering each of the previous results as follows.

\subsubsection{ By  using the $\mu\to e\gamma$ process }

\begin{figure}[H]
	\centering
	\subfloat[]{
		\includegraphics[width=0.5\textwidth]{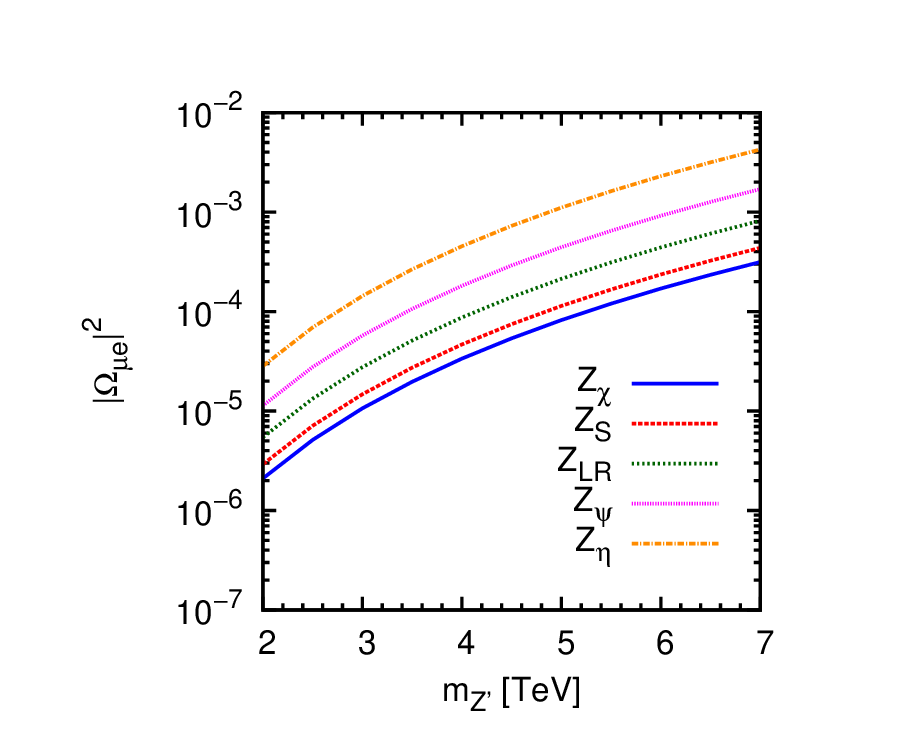}}
	\subfloat[]{
		\includegraphics[width=0.5\textwidth]{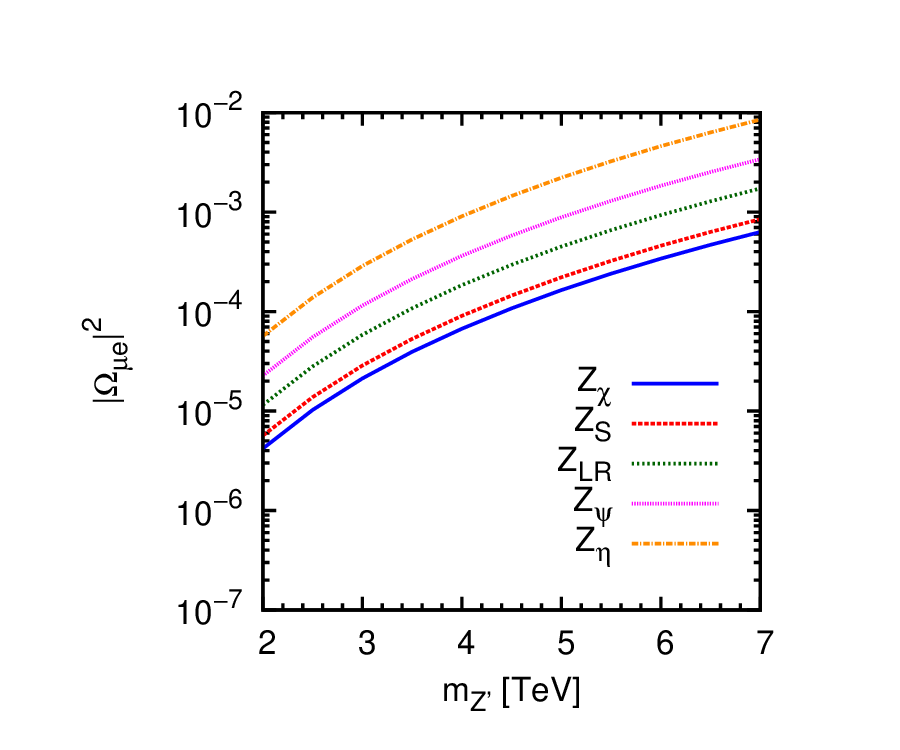}}\hspace{1cm}
	\subfloat[]{
		\includegraphics[width=0.5\textwidth]{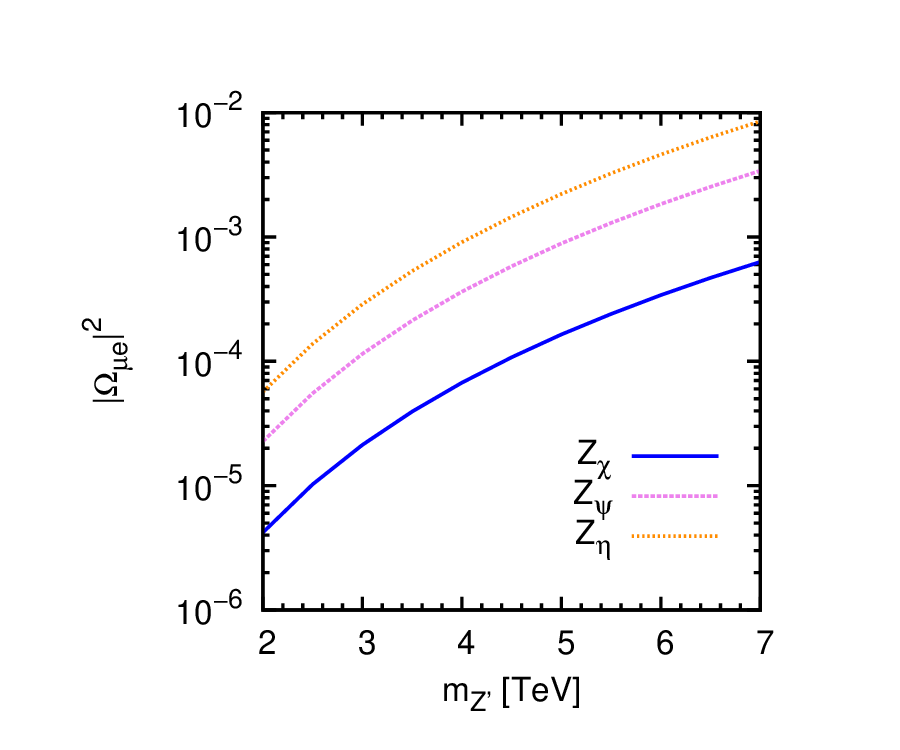}}	
	\caption{Bounds for the $Z^\prime\mu e$ coupling  as a function of $m_{Z^\prime}$, obtained through the  $\mu \to e \gamma$ decay. In (a), vector-like coupling case. In (b),  maximal parity violation case. In (c), the complete case.}
	\label{fig:mueomegaa}
\end{figure}

To numerically constrain the couplings,  we  use   Eq. (\ref{bramu}) along with the experimental bound:  Br$(\mu\to e\gamma)_{\tiny\mathrm{Exp}}<4.2\times 10^{-13}$ with 90\% C.L~\cite{pdg2022}. Also, we take the following experimental values of the masses of the particles involved:  $m_\mu=0.105658$ GeV and $m_e=0.00051099$ GeV. In addition, we use  the most recent experimental limits established for  the mass of the various $Z^\prime$ gauge bosons, which, most of them, are   larger than 3.9 TeV~\cite{aabo,siru}. However, the mass of $Z_{LR}$  boson predicted by the left-right symmetric model is  larger than 1.162 TeV~\cite{Faguila}. In this way,  in next section we propose an analysis for the couplings between 2 and 7 TeV for $m_{Z^\prime}$. For the study, we  consider three different cases:

a) Vector-like coupling. For this case, we have that	$\Omega_{L\mu e}=\Omega_{R\mu e}\equiv\Omega_{\mu e}$.
Therefore, the bound can be expressed as
\begin{align}
	|\Omega_{\mu e}|^2<&\frac{4096 \pi^4}{\alpha}\frac{\Gamma_\mu}{m_\mu}\frac{4.2\times 10^{-13}}{g_2^2 \big[|F_1 (Q_L^e-Q_R^e)+F_2 Q_L^e|^2+|F_1(Q_R^e-Q_L^e)+F_2Q_R^{e}|^2\big]},
	\label{omegamue1}
\end{align}
where we have taken the lower bound of Br$(\mu\to e\gamma)_{\tiny\mathrm{exp}}$ in accordance to Eq.(\ref{bramu}). In Fig. \ref{fig:mueomegaa} (a) it is shown the maxima values of $|\Omega_{\mu e}|^2$ as a function of  $m_{Z^\prime}$ for the different models. As it can  be observed, the most restrictive bound corresponds to the $Z_\chi$ boson, while the less restrictive  corresponds to the $Z_\eta$ boson, along the mass interval studied.

b) Maximal parity violation. For this case, we take $\Omega_{L\mu e}=0$. Thus the corresponding bounding can be written as
\begin{align}
	|\Omega_{R\mu e}|^2<&\frac{4096 \pi^4}{\alpha}\frac{\Gamma_\mu}{m_\mu}\frac{4.2\times 10^{-13}}{g_2^2 \big[|F_1 (Q_L^e-Q_R^e)+F_2 Q_R^e|^2\big]}.
	\label{omegamue2}
\end{align}
In Fig. \ref{fig:mueomegaa} (b) it can be appreciated the behavior of maxima of $|\Omega_{R\mu e}|^2$ as function of $m_{Z^\prime}$ for the different  models under consideration. As before, the most restrictive case corresponds precisely to the  $Z_{\chi}$ boson,  and the  less restrictive case is for the $Z_\eta$ boson.

c) Complete case. This particular case is only achieved for the three different models that share the same absolute value of the left and right chiral charges $|Q_{L}^e|=|Q_{R}^e|$, namely, the
$Z_{\chi}, Z_{\psi}$ and $Z_\eta$ bosons. For this case, we have
\begin{align}
	|\Omega_{L\mu e}|^2+|\Omega_{R\mu e}|^2<&\frac{4096 \pi^4}{\alpha}\frac{\Gamma_\mu}{m_\mu}\frac{4.2\times 10^{-13}}{g_2^2|2 F_1+F_2|^2|Q_L^e|^2}.
	\label{omegamue3}
\end{align}
The maxima values allowed for the sum $|\Omega_{L\mu e}|^2+|\Omega_{R\mu e}|^2$ are depicted in Fig. \ref{fig:mueomegaa} (c). As it can be observed, the most restrictive bound is achieved by the $Z_\chi$  boson.
Finally, let us mention that the  resulting growing behavior of   square magnitude of the  couplings in the mass interval under analysis is consistent with the perturbative regime employed to carry out the study of the $Z^\prime$ boson decay.

\subsubsection{By using the $\mu-e$ conversion process}
Let us now use the $\mu-e$ conversion branching ratio in Eq. (\ref{branchingMue}) to analyze the behavior of the couplings. For the study, we consider the  same three cases above discussed:

a) The vector-like coupling. For this case: $\Omega_{L\mu e}=\Omega_{R\mu e}\equiv \Omega_{\mu e}$. Therefore,
\begin{equation}
	\vert \Omega_{\mu e}\vert^2<\mathrm{Br}(\mu^{-}+N\to e^{-}+N)_{\mathrm{exp}}\:\frac{8\pi^2\,m_{Z'}^4\,Z\,\Gamma_{capture}}{g_2^2\alpha^3\,m_\mu^5 Z_{eff}^4|F(q)|^2 Q_W^{'2}}.\label{mue2}
\end{equation}

b) The coupling with maximal parity violation. For this case $\Omega_{L\mu e}=0$. Thus
\begin{equation}
	\vert \Omega_{R\mu e}\vert^2<\mathrm{Br}(\mu^{-}+N\to e^{-}+N)_{\mathrm{exp}}\:\frac{16\pi^2\,m_{Z'}^4\,Z\,\Gamma_{capture}}{g_2^2\alpha^3\,m_\mu^5 Z_{eff}^4|F(q)|^2 Q_W^{'2}}.\label{mue2}
\end{equation}

c) Complete case. Here, the parameter under consideration is $|\Omega_{L\mu e}|^2+|\Omega_{R\mu e}|^2$. However, this case is the same as to the previous one, since it has the same expression in the right hand of the inequality in (\ref{mue2}).

\begin{figure}[H]
	\centering
	\subfloat[]{
		\includegraphics[width=0.5\textwidth]{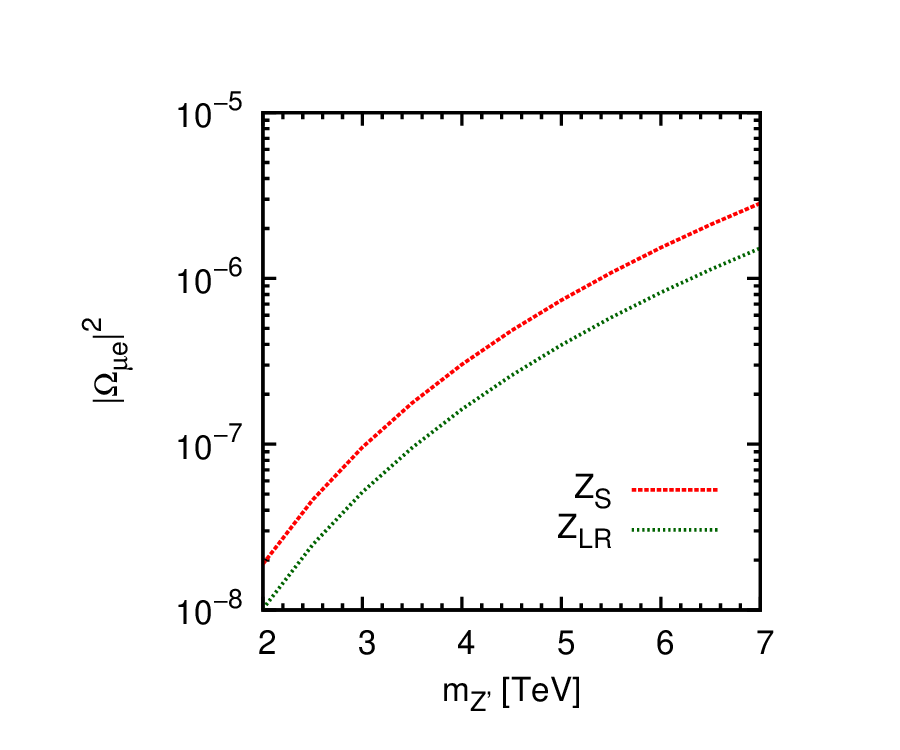}}
	\subfloat[]{
		\includegraphics[width=0.5\textwidth]{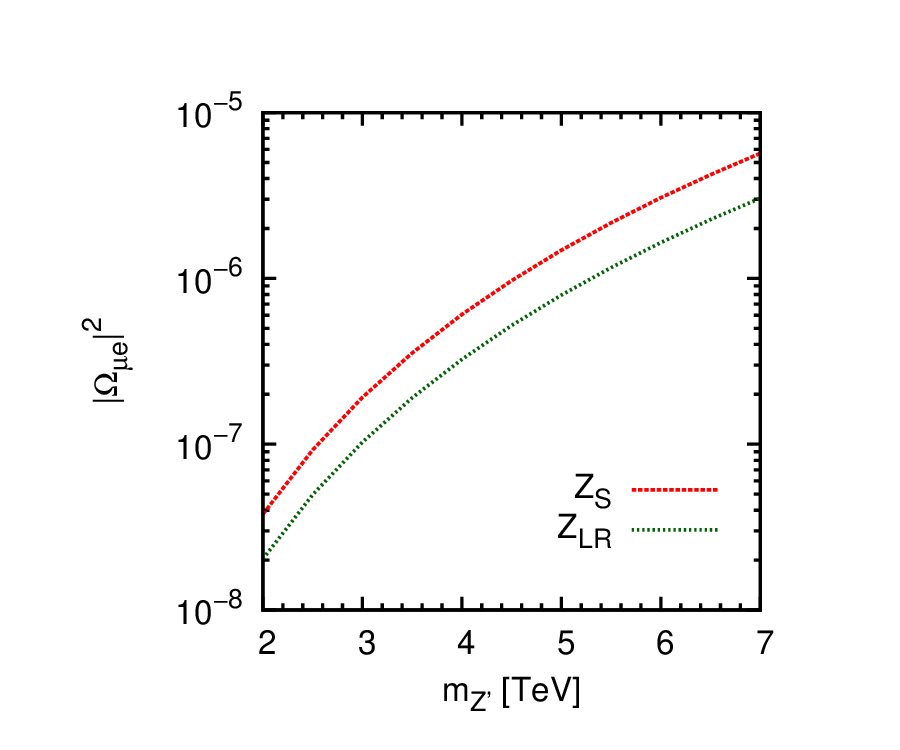}}
	\caption{Bounds for the $|\Omega_{\mu e}|^2$ parameter as function of  $m_{Z^\prime}$ obtained  through the $\mu-e$ conversion process. (a) For the  case of vector-like coupling.  (b) Maximal parity violation coupling and complete cases.}
	\label{OmegaA}
\end{figure}

In Figs. \ref{OmegaA} (a) and (b),   the upper bounds of the couplings in question as a function of $m_{Z^\prime}$ are shown. As it can be observed in these figures, the most restrictive  bounds are provided by the $Z_{LR}$ boson and they result less than $10^{-6}$  almost along the full mass interval.

\subsubsection{By using the $\mu \to e e^{+}e^{-}$ decay}

\begin{figure}[H]
	\centering
	\subfloat[]{
		\includegraphics[width=0.5\textwidth]{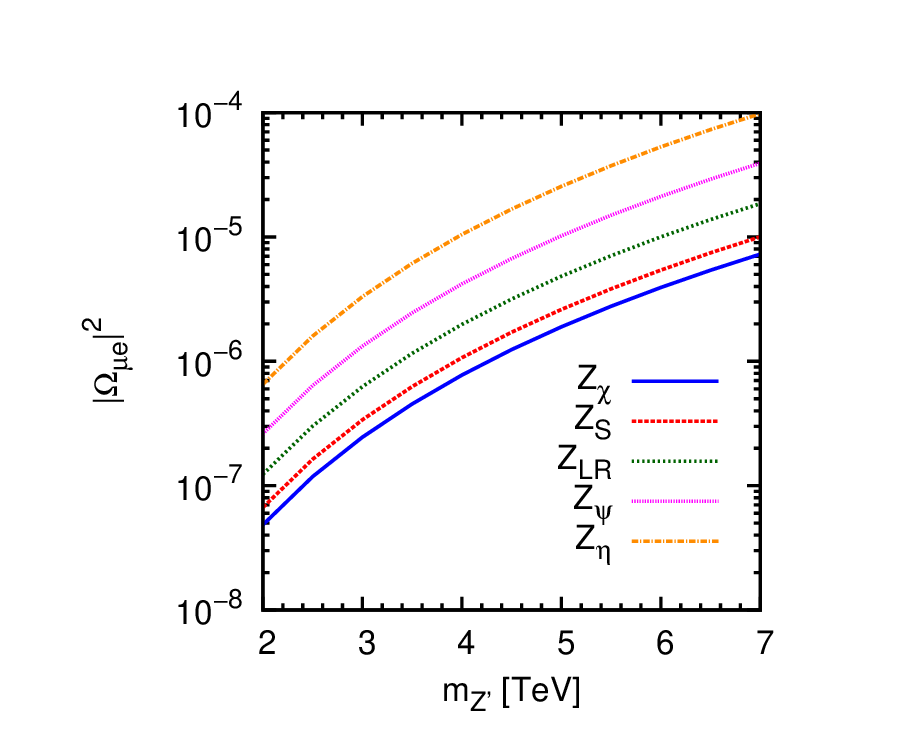}}
	\subfloat[]{
		\includegraphics[width=0.5\textwidth]{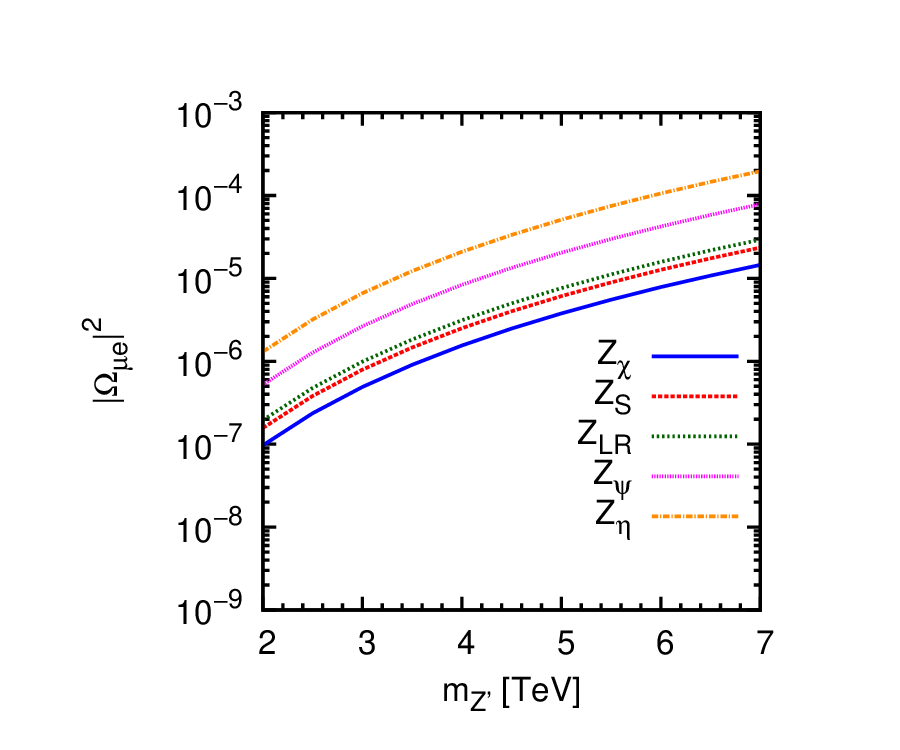}}\hspace{1cm}
	\subfloat[]{
		\includegraphics[width=0.5\textwidth]{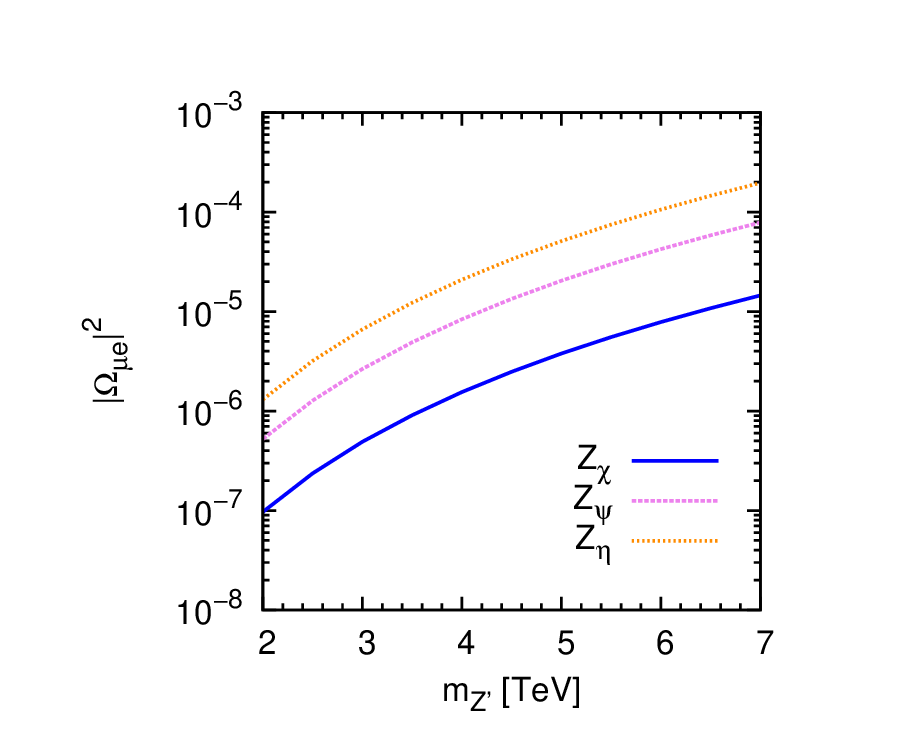}}	
	\caption{Bounds for the $|\Omega_{\mu e}|^2$ parameter as function of  $m_{Z^\prime}$ obtained  through the $\mu\to e e^{+}e^{-}$ decay. In (a), vector-like coupling case. In (b),  maximal parity violation case. In (c), the complete case.}
	\label{muetree}
\end{figure}

Once again, we consider the three cases  previously discussed. To accomplish the task of bounding the couplings,  we use Eq. (\ref{three-decay-bis}) along with the experimental bound  $\mathrm{Br}(\mu \to e e^{+}e^{-})_{\mathrm{exp}}<1.0\times 10^{-12}$~\cite{pdg2022}. The corresponding results are listed below.

a) Vector-like coupling: The corresponding bound is
\begin{align}
	|\Omega_{\mu e}|^2<384\pi^3\frac{\Gamma_\mu}{m_\mu}\frac{\mathrm{Br}(\mu\to ee^+e^-)_{\mathrm{exp}}}{\big[g_2^2({Q_L^e}^2+{Q_R^e}^2)\big]f(\frac{m_{Z^\prime}^2}{m_\mu^2})}.
\end{align}

b) Maximal parity violation coupling:
\begin{align}
	|\Omega_{R\mu e}|^2<384\pi^3\frac{\Gamma_\mu}{m_\mu}\frac{\mathrm{Br}(\mu\to ee^+e^-)_{\mathrm{exp}}}{g_2^2{Q_R^e}^2f(\frac{m_{Z^\prime}^2}{m_\mu^2})}.
\end{align}

c) Complete case:
\begin{align}
	|\Omega_{L\mu e}|^2+|\Omega_{R\mu e}|^2<384\pi^3\frac{\Gamma_\mu}{m_\mu}\frac{\mathrm{Br}(\mu\to ee^+e^ -)_{\mathrm{exp}}}{g_2^2{Q_L^e}^2f(\frac{m_{Z^\prime}^2}{m_\mu^2})}.
\end{align}
In Fig.~\ref{muetree}, it is shown the behavior of  bounds in question. As it can be observed, the most restrictive bounds  for the couplings come from the  $Z_\chi$ boson in the three   cases. On the contrary, the less restrictive bounds are provided by the  $Z_\eta$ boson. In general, the case of vector-like coupling offers the most restrictive bounds. Although, as the bounds arising from the  $Z_S$ and $Z_{LR}$ gauge bosons are concerned, the most restrictive ones  result from the $\mu-e$ conversion process, as it can be appreciated from Figs. \ref{OmegaA} (a) and \ref{muetree} (a).

\section{The $Z^\prime\to \mu e $  decay}\label{decays}

The Feynman rules for this process can be  obtained from the Lagrangian  in~(\ref{1a}). By following the usual procedures, we obtain that the general branching ratio for the lepton flavor violating process $Z^\prime\to l_il_j$
can be written as
\begin{eqnarray}
	\textup{Br}(Z^\prime\to l_i l_j)\approx
	\frac{m_{Z^\prime}}{12\pi\,\Gamma_{Z^\prime}}\bigg(|\Omega_{L l_il_j}|^2+|\Omega_ {R l_il_j}|^2\bigg),
	\label{br}
\end{eqnarray}
where $\Gamma_{Z^\prime}$ is the total decay width of the $Z^\prime$ boson, and it was considered that $m_{\mu}/m_{Z^\prime} \ll 1$.  On account of $\Gamma_{Z^\prime}$, we consider
all possible decay modes that include both flavor-conserving and flavor-violating decays~\cite{aranda2012,arhrib,aranda2011}, that is to say, the decay of $Z^\prime$ to  $\nu_e \bar{\nu_e}, \nu_\mu \bar{\nu_\mu}, \nu_\tau \bar{\nu_\tau}, e\bar{e}, \mu\bar{\mu}, \tau\bar{\tau}, u\bar{u}, c\bar{c}, t\bar{t}, d\bar{d}, s\bar{s}, b\bar{b}, \bar{u}c+u\bar{c}, \bar{t}c+t\bar{c}, \bar{\tau}\mu+\tau\bar{\mu}, \bar{\tau}e+\tau\bar{e}$, and $\bar{\mu} e+\bar{e}\mu$. Notice that the total decay width $\Gamma_{Z^\prime}$ is model dependent, since each  decay mode of this particle is also model dependent.
Previously, we have analyzed three different cases to bounding the couplings, through the three process above discussed, which will be used in the numerical analysis of  the $Z^\prime\to\mu e$ decay. For the discussion, we restrict ourselves to the vector-like coupling case, which gives the most restrictive bounds for the decay in question. As matter of fact, the others cases provide bounds of the same order of magnitude for each $m_{Z^\prime}$, along the mass interval. Due to the values of the chiral charges (which make the branching ratio in (\ref{br}) a non-trivial function of these charges), and the $\Omega$ couplings,  the bounds  for the branching ratio can change up to three orders of magnitude, as it is discussed below.

\begin{figure}[H]
	\centering
	\subfloat[]{
		\includegraphics[width=0.5\textwidth]{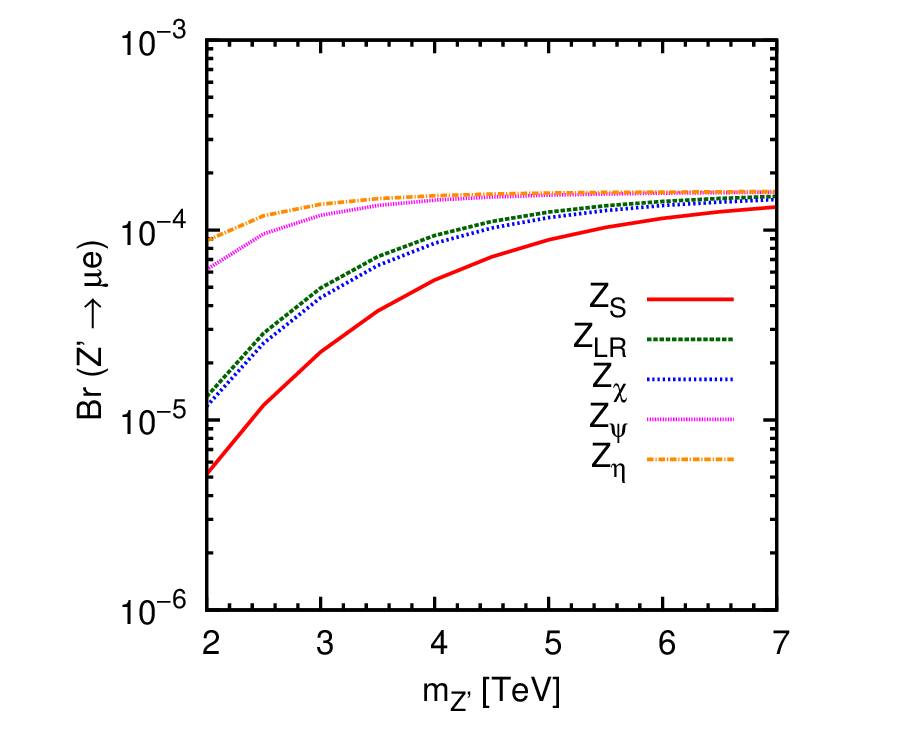}}
	\subfloat[]{
		\includegraphics[width=0.5\textwidth]{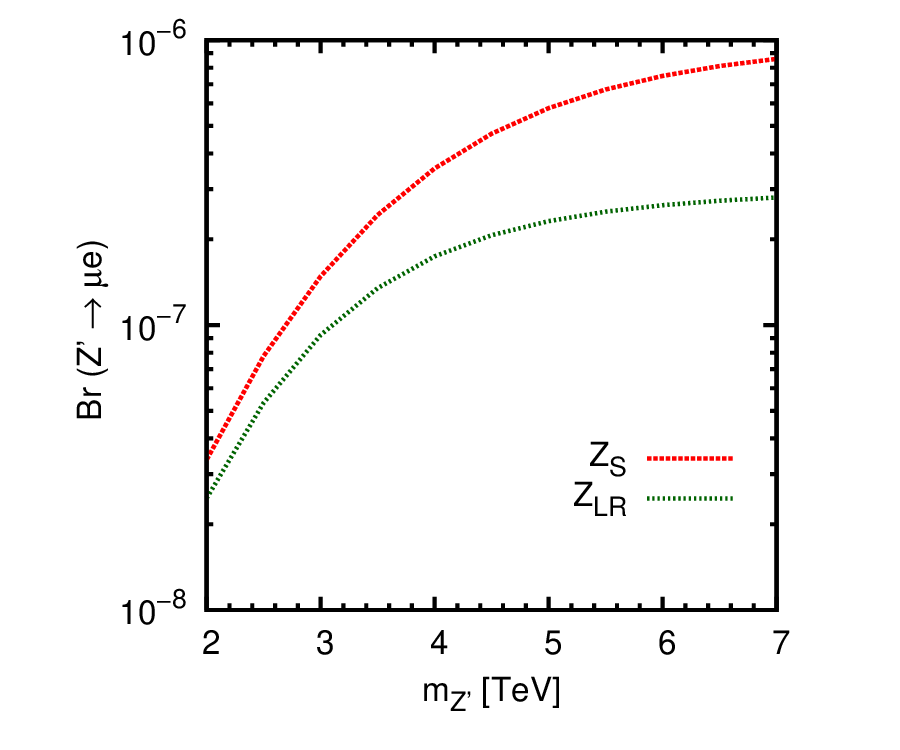}}\hspace{1cm}
	\subfloat[]{
		\includegraphics[width=0.5\textwidth]{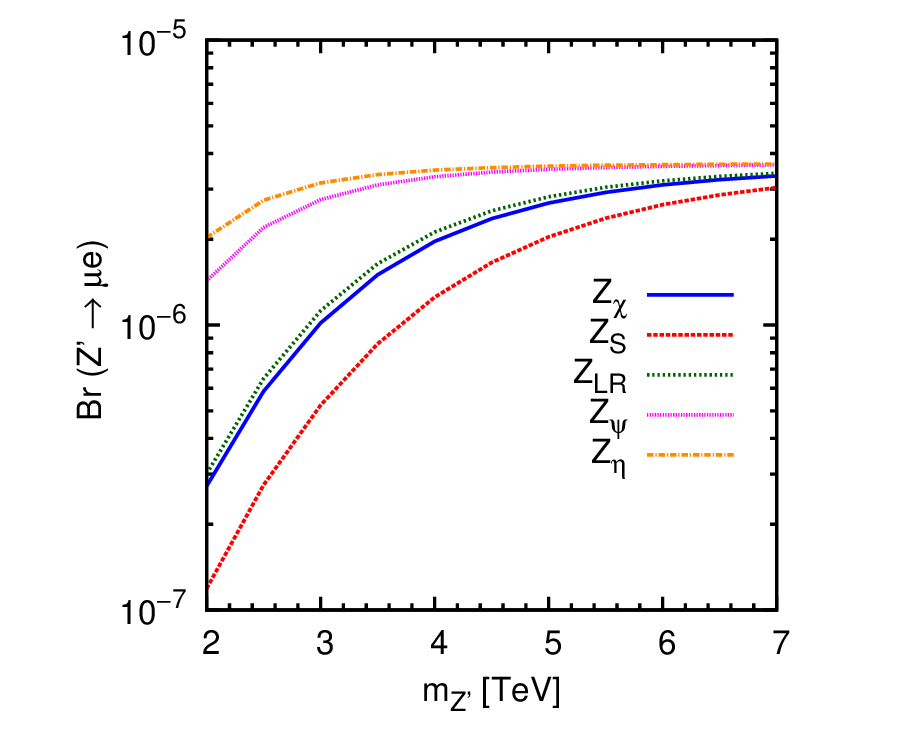}}
	\caption{The branching ratios for the LFV decay of $Z^\prime\to \mu e$ as a function of $m_{Z^\prime}$. The branching ratio was obtained by means of:  (a) The $\mu \to e \gamma$ decay.  (b) The  $\mu-e$ conversion process in nuclei. (c) The $\mu \to e e^{+}e^{-}$ decay.}
	\label{fig4}
\end{figure}

In Fig.~\ref{fig4} (a) it is shown the bounds of the branching ratio  for the  $Z^\prime\to \mu e$ decay, calculated by using the $\mu\to e \gamma$ decay.  As we can see,  the $Z_\eta$ boson provides the less restrictive bound for the decay, with values less than $10^{-4}$ almost over the entire interval [2,7] TeV. The most suppressed bound corresponds to the $Z_S$ boson, which gives values for $\mathrm{Br}(Z^\prime\to \mu e)< 10^{-5}$ for masses  below  5 TeV,  reaching the  value of  $1.32\times 10^{-4}$ at $m_{Z^\prime}=$ 7 TeV. For the rest of  the  models, the corresponding branching ratios vary between $10^{-5}$ and $10^{-4}$.
%

Let us now discuss the results of bounds for the branching ratio by using the predictions for the $\Omega_{\mu e}$ coupling coming from the $\mu-e$ conversion process. In Fig.~\ref{fig4} (b),  the behavior of   such bounds is shown as a function of $m_{Z^\prime}$. As it can be observed, the resulting bounds for the $Z_S$  and $Z_{LR}$ bosons are more restrictive than the  corresponding ones obtained through the $\mu\to e\gamma$ decay   (shown in Fig. \ref{fig4} (a)), since the less restrictive result for the branching ratio is  $<10^{-6}$, even in the most conservative scenario, which is obtained with the $Z_S$ boson.

As far as the bounds for $\mathrm{Br}(Z^\prime\to \mu e)$ resulting from the $\mu \to e e^{+}e^{-}$ decay is concerned, the results are shown in Fig. \ref{fig4} (c). The numerical values for the bounds show that the corresponding branching ratios are more suppressed  than the obtained through the $\mu\to e\gamma$ decay by two order of magnitude for each $Z^\prime$  boson. Nevertheless, the only   non-vanishing  branching ratios (for $Z_S$ and $Z_{LR}$  bosons),  obtained by means of the $\mu-e$ conversion process, result more suppressed along the mass interval analyzed than the  obtained by using the  $\mu\to e\gamma$ decay or the $\mu \to e e^{+}e^{-}$ decay.

\begin{figure}[H]
	\centering
	\subfloat
		{\includegraphics[scale=.5]{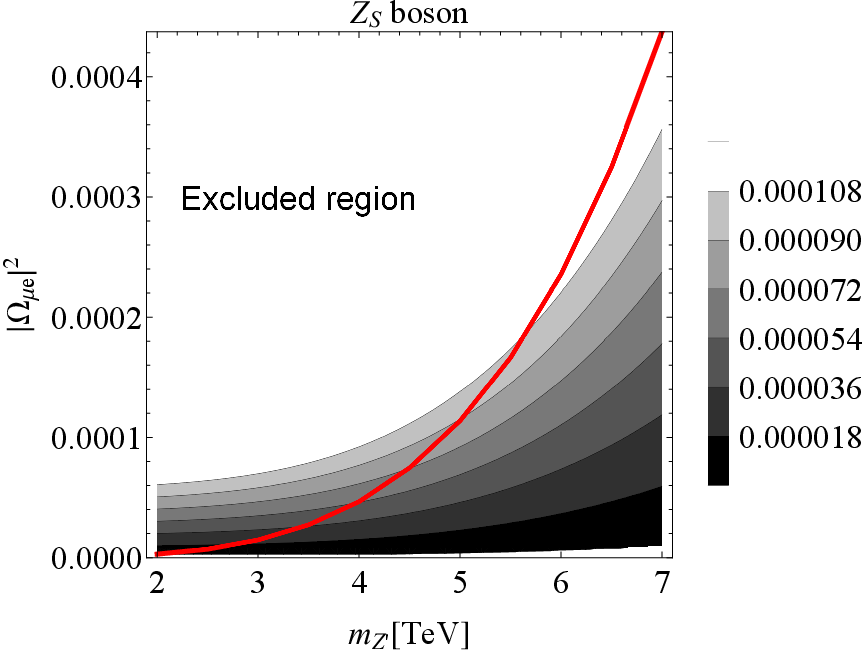}}
		\subfloat[]{
			\includegraphics[scale=.5]{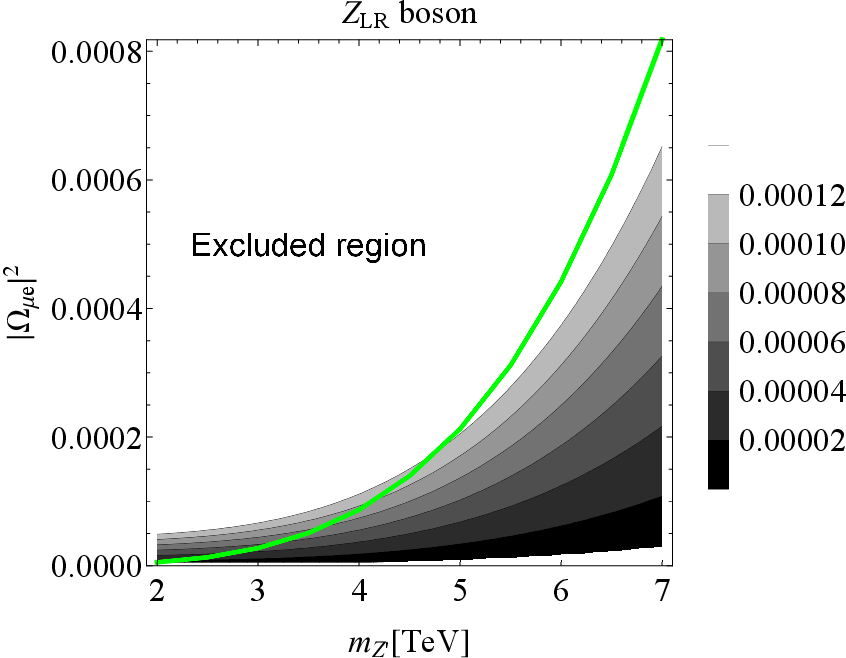}}\hspace{1cm}
		\subfloat[]{
			\includegraphics[scale=.5]{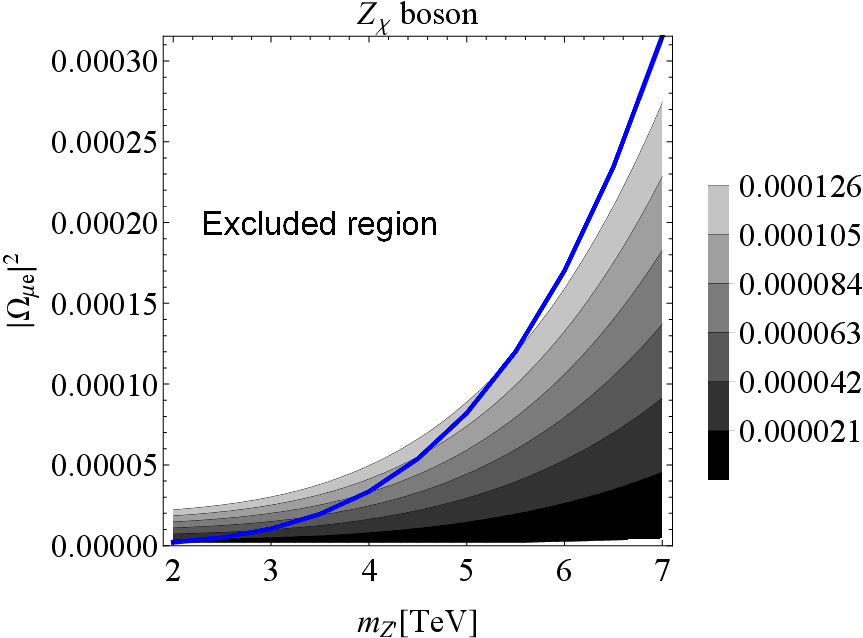}}
		\subfloat[]{
			\includegraphics[scale=.5]{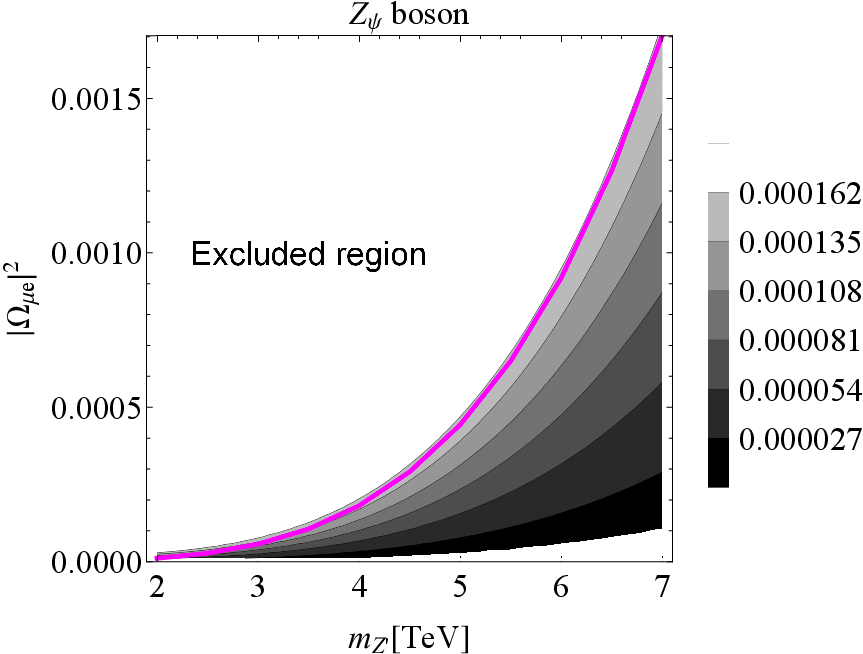}}\hspace{1cm}
		\subfloat[]{
			\includegraphics[width=.5\textwidth]{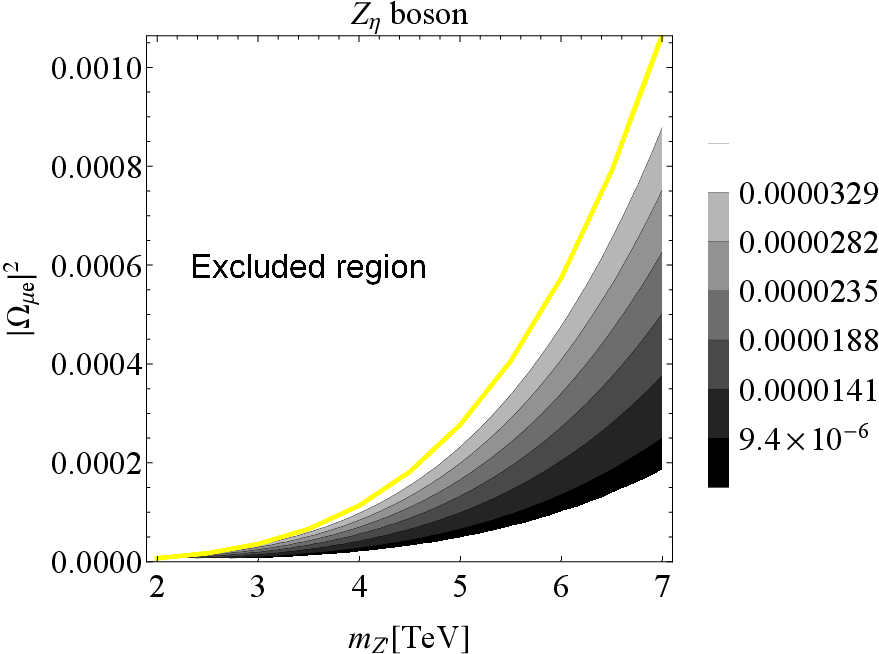}}
		\caption{ Contour plots for the Br$(Z^\prime\to \mu e)$. The graphs were obtained  by using  the $\mu \to e \gamma$ decay  process. In (a) for the $Z_S$ boson,   in (b) for the $Z_{LR}$ boson, in (c) for the $Z_{\chi}$ boson,  in (d) for the $Z_{\psi}$ boson, and in (e) for the $Z_{\eta}$ boson. In all the cases the thick line in color  separates the excluded from the allowed region.}
		\label{contours1}
	\end{figure}

\begin{figure}[H]
		\centering
		\subfloat[]{
			\includegraphics[scale=.5]{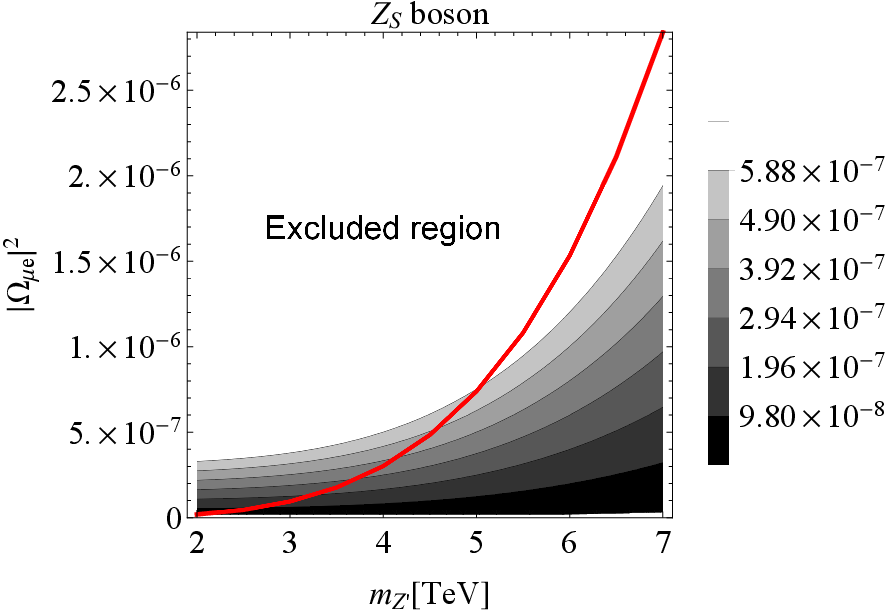}}
		\subfloat[]{
			\includegraphics[scale=.5]{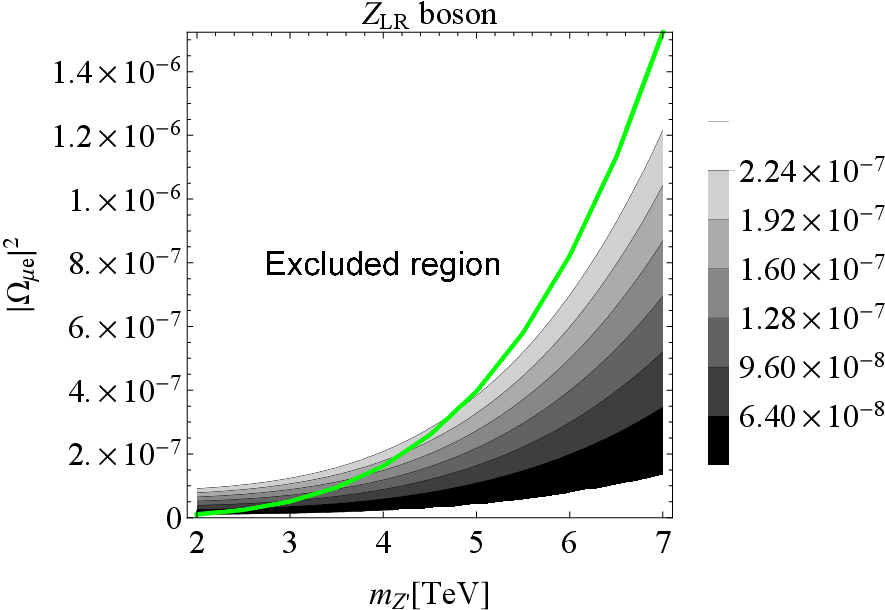}}
		\caption{ Contour plots for the Br$(Z^\prime\to \mu e)$. The graphs were obtained  by using  the $\mu-e$  conversion process. In (a) for the $Z_S$ boson and  in (b) for the $Z_{LR}$ boson. In both cases the thick line in color   separates the excluded  from the allowed region.}
		\label{contours2}
	\end{figure}

\begin{figure}[H]
		\centering
		\subfloat
			{\includegraphics[scale=.5]{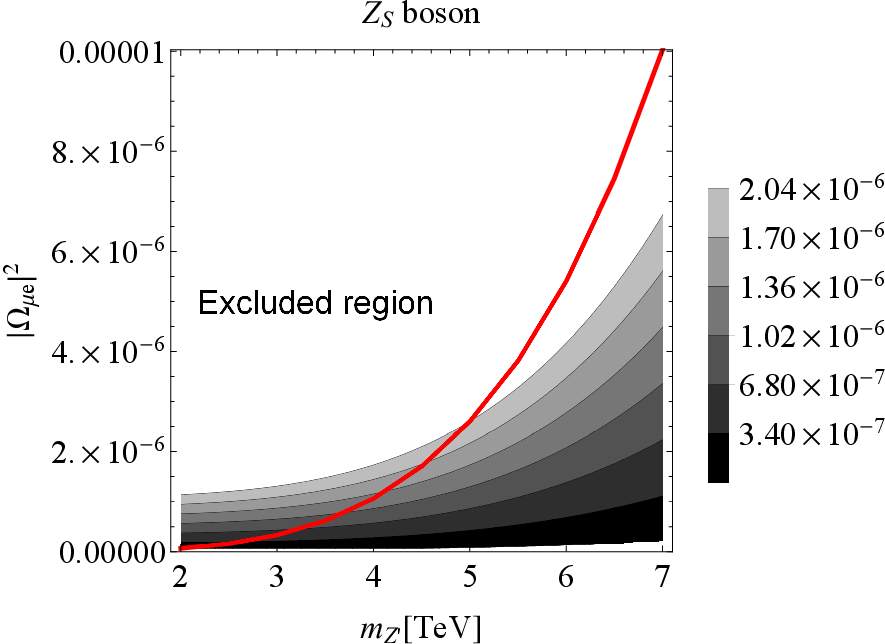}}
			\subfloat[]{
				\includegraphics[scale=.5]{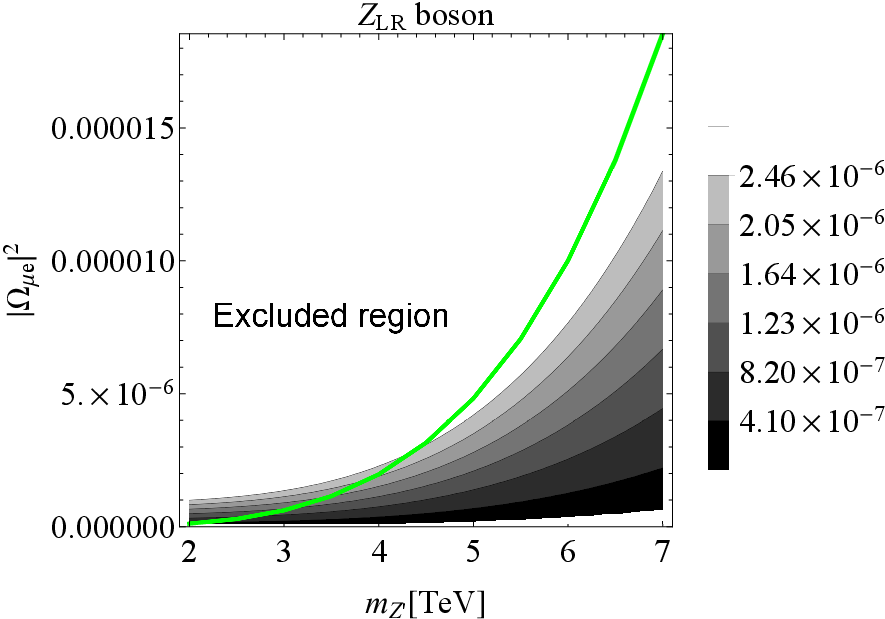}}\hspace{1cm}
			\subfloat[]{
				\includegraphics[scale=.5]{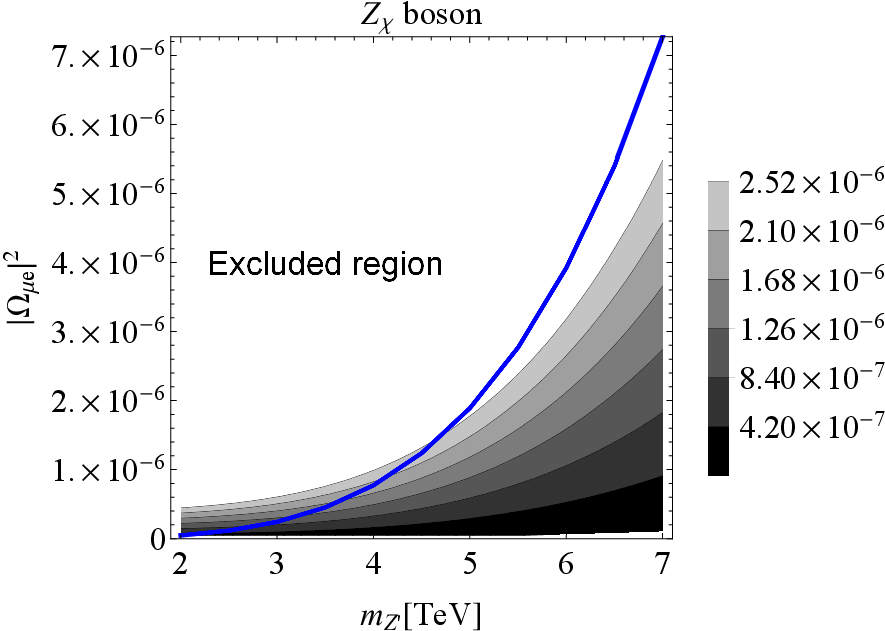}}
			\subfloat[]{
				\includegraphics[scale=.5]{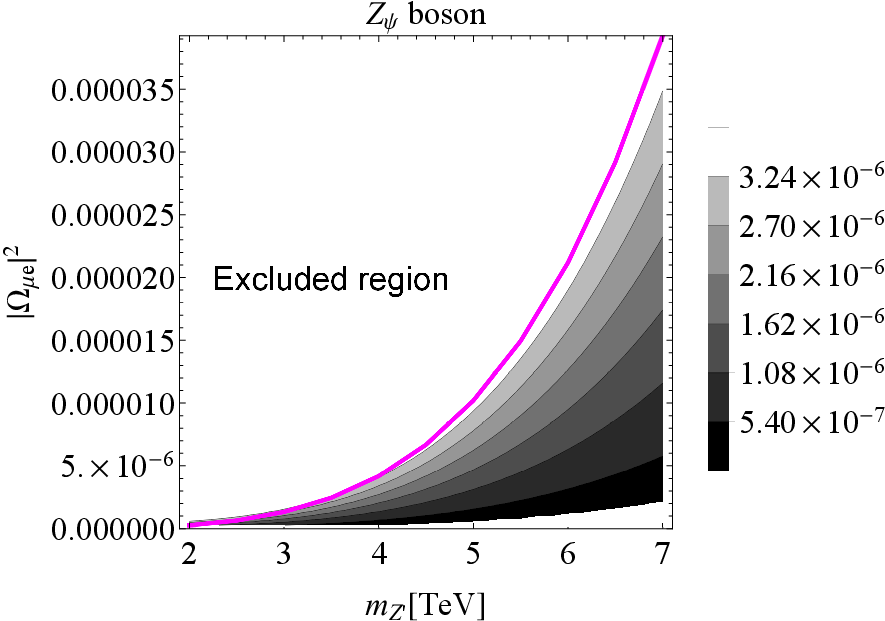}}\hspace{1cm}
			\subfloat[]{
				\includegraphics[scale=.5]{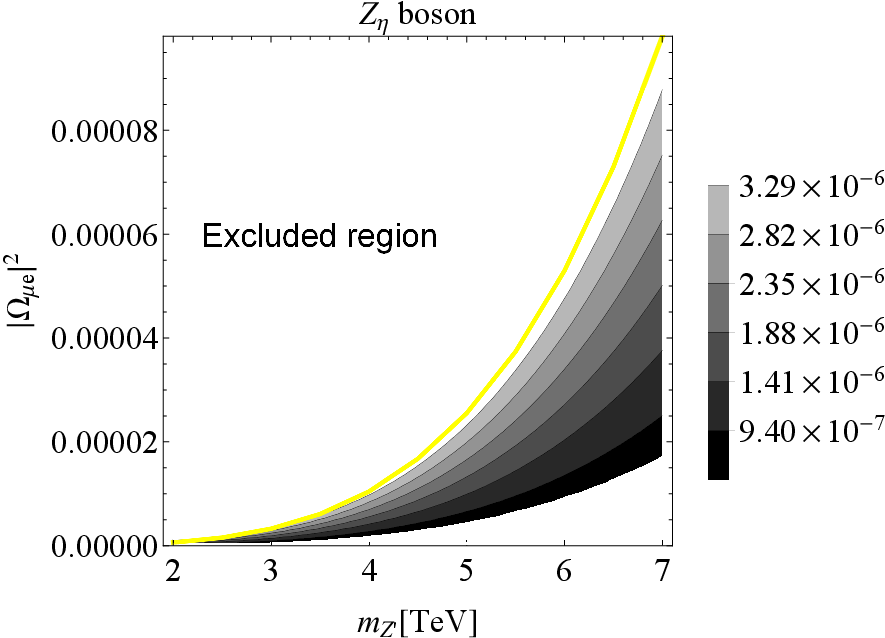}}
			\caption{ Contour plots for the Br$(Z^\prime\to \mu e)$. The graphs were obtained  by using  the $\mu \to e e^{+}e^{-}$ decay  process. In (a) for the $Z_S$ boson,   in (b) for the $Z_{LR}$ boson, in (c) for the $Z_{\chi}$ boson,  in (d) for the $Z_{\psi}$ boson, and in (e) for the $Z_{\eta}$ boson. In all the cases the thick line in color  separates the excluded from the allowed region.}
			\label{contours3}
		\end{figure}

Complementarily, we also present contour plots for the Br$(Z^\prime\to \mu e)$ as function of $m_{Z\prime}$ and $|\Omega|^2$ (see Figs. \ref{contours1}, \ref{contours2} and \ref{contours3}) only for the vector-like  case, since it provides the most stringent predictions.  In these plots (one for each model and  bounding approach) it is shown the excluded regions
for specific branching ratios.
	
\section{Conclusions}\label{conclu}
The $Z^\prime$ gauge boson is   proposed  in several extensions of the SM that induces FCNC at the tree level. This particle is under quest by the  CMS and ATLAS experimental groups at LHC who have established experimental bounds on its mass. The existence of this particle could explain physical processes, such as the flavor-violating decays in the sector of leptons that  are very suppressed in the SM, whose numerical values  are not   compatible with the respective experimental results. In this work,  we have studied the flavor-violating branching ratio of decay of the $Z^\prime$ gauge  boson into $\mu e$  leptons, using  different extended models and three different ways of bounding the decay. By resorting to the $\mu\to e\gamma$ process, we found bounds for the  $\mathrm{Br}(Z^\prime\to\mu e)$ as function of  $m_{Z^\prime}$; depending on the model, {the values go from $10^{-6}$ to $10^{-4}$, along  the mass interval analyzed.
If instead,  the $\mu-e$ conversion process is used, we obtain more restrictive bounds for the branching ratio of the $Z^{\prime}\to \mu e$ decay. In specific, the values of the branching ratio  go from $10^{-8}$ to $10^{-6}$.
However, due to the  property    $Q_L^q+Q_R^q=0$ of the chiral charges of quarks the only non-vanishing branching ratios for this process correspond to the $Z_S$ and $Z_{LR}$ bosons.
For the case of the tree-body decay: $\mu \to e e^{+}e^{-}$,  the resulting bounds for $\mathrm{Br}(Z^\prime\to\mu e)$ are more suppressed than the obtained in the cases above discussed, except for that related to the $Z_S$ and $Z_{LR}$ bosons.  In fact, in this case the bounds of the branching ratio go from $10^{-7}$ to $10^{-6}$.
If we compare these bounds with the current experimental  bound for the SM  decay $\mathrm{Br}(Z\to\mu e)<7.5\times 10^{-7}$, the bounds for the  $\mathrm{Br}(Z^\prime\to\mu e)$ result less restrictive or at least comparable.

\acknowledgements

This work has been partially supported by CIC-UMSNH and CONAHCYT, M\'exico.


\section*{Appendix}
\noindent Form Factors:
\begin{small}
	\begin{eqnarray}
		F_1 &=&6+x_\mu^2+\frac{3(3x_\mu^2-1)}{x_\mu^2(x_\mu^2-1)}\textup{Log}(x_\mu^2)+6\frac{\big(1-4x_\mu^2\big)^{1/2}}{x_\mu^2}\textup{Log}\bigg(\frac{1+\big(1-4x_\mu^2\big)^{1/2}}{2x_\mu}\bigg),
		\nonumber
	\end{eqnarray}
\end{small}

\begin{eqnarray}
	 F_2&=&8+2\Bigg[-\frac{4}{x_\mu^2}-\frac{3x_\mu^2-1}{x_\mu^4(x_\mu^2-1)}\textup{Log}(x_\mu^2)+\textup{Log}\bigg(\frac{1+\big(1-4x_\mu^2\big)^{1/2}}{2}\bigg)^2\nonumber\\
	 &+&2\bigg[\bigg(\frac{x_\mu^2-1}{x_\mu^2}\bigg)^2\textup{Log}\bigg(\frac{1}{1-x_\mu^2}\bigg)-\frac{\big(1-4x_\mu^2\big)^{1/2}}{x_\mu^4}\textup{Log}\bigg(\frac{1+\big(1-4x_\mu^2\big)^{1/2}}{2x_\mu}\bigg)\nonumber\\
	 &+&\textup{Polylog}\bigg(2,1-\frac{1}{x_\mu^2}\bigg)+\textup{Polylog}\bigg(2,\frac{2x_\mu^2}{1+\big(1-4x_\mu^2\big)^{1/2}}\bigg)\nonumber\\
	&-&\textup{Polylog}\bigg(2,\frac{2x_\mu^2}{-1+2x_\mu^2+\big(1-4x_\mu^2\big)^{1/2}}\bigg)\bigg]\Bigg],
	\nonumber
\end{eqnarray}
where $x_\mu=\frac{m_\mu}{m_{Z^\prime}}$.


\begin{thebibliography}{99}
	
	\bibitem{SMqsectorsup} For instance, see G. Eilam, J. L. Hewett, and A. Soni, \href{https://doi.org/10.1103/PhysRevD.44.1473}{{\textcolor{blue}{Phys. Rev. D \textbf{44}, 1473 (1991)}}}; \href{https://doi.org/10.1103/PhysRevD.59.039901}{{\textcolor{blue}{\textbf{59}, 039901(E) (1998)}}}; N. G. Deshpande, B. Margolis, and H. D. Trottier, \href{https://doi.org/10.1103/PhysRevD.45.178}{{\textcolor{blue}{Phys. Rev. D \textbf{45}, 178 (1992)}}}; B. Mele, S. Petrarca, and A Soddu, \href{https://doi.org/10.1016/S0370-2693(98)00822-3}{{\textcolor{blue}{Phys. Lett. \textbf{B 435}, 401 (1998)}}}; A. Cordero-Cid, J. M. Hern\'andez, G. Tavares-Velasco, and J. J. Toscano, \href{https://doi.org/10.1103/PhysRevD.73.094005}{{\textcolor{blue}{Phys. Rev. D \textbf{73}, 094005 (2006)}}}; G. Eilam, M. Frank, and I. Turan, \href{https://doi.org/10.1103/PhysRevD.73.053011}{{\textcolor{blue}{Phys. Rev. D \textbf{73}, 053011 (2006)}}}; \href{https://doi.org/10.1103/PhysRevD.74.035012}{\textcolor{blue}{\textbf{74}, 035012 (2006)}}, M. Artuso \textit{et al}, \href{https://doi.org/10.1140/epjc/s10052-008-0716-1}{\textcolor{blue}{Eur. Phys. J. C \textbf{57}, 309-492 (2008)}}.
	
	
	\bibitem{becker} R. Becker-Szendy, C. B. Bratton, D. Casper, S. T. Dye, W. Gajewski, M. Goldhaber \textit{et al}.,   \href{https://doi.org/10.1103/PhysRevD.46.3720}{\textcolor{blue}{Phys.  Rev. D \textbf{46}, 3720 (1992)}};
	Y. Fukuda \textit{et al}., \href{https://doi.org/10.1016/0370-2693(94)91420-6}{\textcolor{blue}{Phys.  Lett. \textbf{B335}, 237 (1994)}}; \href{https://doi.org/10.1103/PhysRevLett.81.1562}{\textcolor{blue}{Phys.  Rev. Lett.  \textbf{81}, 1562 (1998)}}; H. Sobel, \href{https://doi.org/10.1016/S0920-5632(00)00932-4}{\textcolor{blue}{Nucl. Phys.  \textbf{B}, Proc. Suppl. \textbf{91}, 127 (2001)}}; M. Ambrossio \textit{et al}., \href{https://doi.org/10.1016/S0370-2693(03)00806-2}{\textcolor{blue}{Phys.  Lett.  \textbf{B566}, 35 (2003)}}; Y. Ashie \textit{et al}., \href{https://doi.org/10.1103/PhysRevD.71.112005}{\textcolor{blue}{Phys.  Rev. D \textbf{71}, 112005 (2005)}}; W. W. M. Allison, G. J. Alner, D. S. Ayres, G. D. Barr, W. L. Barrett,  P. M. Border,  \textit{et al}., \href{https://doi.org/10.1103/PhysRevD.72.052005}{\textcolor{blue}{  Phys.  Rev. D \textbf{72}, 052005 (2005)}}; P. Adamson \textit{et al}., \href{https://doi.org/10.1103/PhysRevD.73.072002}{\textcolor{blue}{Phys. Rev. D \textbf{73}, 072002 (2006)}}.
	
	\bibitem{apollonio} M. Apollonio \textit{et al}., \href{https://doi.org/10.1140/epjc/s2002-01127-9}{\textcolor{blue}{Eur. Phys. J. C \textbf{27}, 331 (2003)}}; S. N. Ahmed \textit{et al}., \href{https://doi.org/10.1103/PhysRevLett.92.181301}{\textcolor{blue}{Phys.  Rev. Lett.  \textbf{92}, 181301 (2004)}};  M. B. Smy \textit{et al}., \href{https://doi.org/10.1103/PhysRevD.69.011104}{\textcolor{blue}{Phys.  Rev. D \textbf{69}, 011104 (2004)}}; E. Aliu \textit{et al}., \href{https://doi.org/10.1103/PhysRevLett.94.081802}{\textcolor{blue}{Phys. Rev. Lett.  \textbf{94}, 081802 (2005)}}. \bibitem{fukuda} Y. Fukuda, \textit{et al}., \href{https://doi.org/10.1103/PhysRevLett.82.2644}{\textcolor{blue}{Phys. Rev. Lett. \textbf{82}, 2644 (1999)}}; Q. R. Ahmad, \textit{et al}., \href{https://doi.org/10.1103/PhysRevLett.89.011301}{\textcolor{blue}{Phys. Rev. Lett. \textbf{89}, 011301 (2002)}};  K. Eguchi, \textit{et al}., \href{https://doi.org/10.1103/PhysRevLett.90.021802}{\textcolor{blue}{Phys. Rev. Lett. \textbf{90}, 021802 (2003)}}.
	
	\bibitem{cheng-li} T-P Cheng, L-F Li, \textit{Gauge Theory of Elementary Particle Physics},  Clarendon Press, Oxford (1984).
	
	\bibitem{pdg2022} R.L. Workman et al. (Particle Data Group), \href{https://doi.org/10.1093/ptep/ptac097}{\textcolor{blue}{Prog. Theor. Exp. Phys. 2022, 083C01 (2022)}}.
	
	
	\bibitem{mendez-mir-otros} A. M\'endez, Ll. M. Mir, \href{https://doi.org/10.1103/PhysRevD.40.251}{\textcolor{blue}{Phys. Rev. D \textbf{40}, 251 (1989)}}; S. Nussinov, R. D. Peccei,and X. M. Zhang, \href{https://doi.org/10.1103/PhysRevD.63.016003}{\textcolor{blue}{Phys. Rev. D \textbf{63}, 016003 (2000)}}; D, Del\'epine and F. Vissani, \href{https://doi.org/10.1016/S0370-2693(01)01254-0}{\textcolor{blue}{Phys. Lett.  \textbf{B522}, 95 (2001)}}; E. O. Iltan and I. Turan, \href{https://doi.org/10.1103/PhysRevD.65.013001}{\textcolor{blue}{Phys. Rev. D  \textbf{65}, 013001 (2001)}}; A. Flores-Tlalpa, J. M. Hern\'andez, G. Tavares-Velasco, and J. J. Toscano, \href{https://doi.org/10.1103/PhysRevD.65.073010
	}{\textcolor{blue}{Phys. Rev. D  \textbf{65}, 073010 (2002)}}; M. A. Perez, G. Tavares-Velasco, and J. J. Toscano, \href{https://doi.org/10.1142/S0217751X04017100
	}{\textcolor{blue}{Int. J. Mod. Phys. A  \textbf{19}, 159 (2004)}}; L. Calibbi, X. Marcano, and J. Roy, \href{https://doi.org/10.1140/epjc/s10052-021-09777-3
	}{\textcolor{blue}{Eur. Phys. J. C \textbf{81}, 1054 (2021)}}.

	\bibitem{robinet1} R. W. Robinett  and J. L. Rosner, \href{https://doi.org/10.1103/PhysRevD.26.2396
	}{\textcolor{blue}{Phys. Rev. D \textbf{26}, 2396 (1982)}};
	R. W. Robinett, \href{https://doi.org/10.1103/PhysRevD.26.2388
	}{\textcolor{blue}{Phys. Rev. D26, 2388 (1982)}}.
	

	
	
\bibitem{lang2} P. Langacker and M. Luo, \href{https://doi.org/10.1103/PhysRevD.45.278}{\textcolor{blue}{Phys. Rev. D \textbf{45}, 278 (1992)}}.

\bibitem{Zwin} F. Zwirner, \href{https://doi.org/10.1142/S0217751X88000035}{\textcolor{blue}{Int. J. Mod. Phys. A3, 49 (1988)}}.

\bibitem{JRizo} J.L. Hewett and T.G. Rizzo,
\href{https://doi.org/10.1016/0370-1573(89)90071-9}{\textcolor{blue}{Phys. Rep. \textbf{183}, 193 (1989)}}.
	
\bibitem{perez-soriano} M. A. Perez and M. A. Soriano, \href{https://doi.org/10.1103/PhysRevD.46.284}{\textcolor{blue}{Phys. Rev. D \textbf{46}, 284 (1992)}}.
	
	\bibitem{framton}P.H. Frampton, \href{https://doi.org/10.1103/PhysRevLett.69.2889
	}{\textcolor{blue}{Phys. Rev. Lett. \textbf{69}, 2889 (1992)}}.
	
	\bibitem{leike} A. Leike, \href{https://doi.org/10.1016/S0370-1573(98)00133-1
	}{\textcolor{blue}{Phys. Rept.  \textbf{317}, 143 (1999)}}.
	
	\bibitem{arhrib} A. Arhrib,  K. Cheung, C-W Chiang, and T-C Yuan, \href{https://doi.org/10.1103/PhysRevD.73.075015
	}{\textcolor{blue}{Phys. Rev. D \textbf{73}, 075015 (2006)}}.
	
	\bibitem{lan2009} Paul Langacker, \href{https://doi.org/10.1103/RevModPhys.81.1199
	}{\textcolor{blue}{Rev. Mod. Phys. \textbf{81}, 1199 (2009)}}.
	
	\bibitem{aabo} M. Aaboud \textit{et al}. (ATLAS Collaboration), \href{https://doi.org/10.1016/j.physletb.2016.08.055}{\textcolor{blue}{Phys. Lett. \textbf{B761}, 372 (2016)}};
	
	
	\bibitem{siru} A. M. Sirunyan \textit{et al}. (CMS Collaboration), \href{https://doi.org/10.1007/JHEP06(2018)120}{\textcolor{blue}{JHEP \textbf{06}, 120 (2018)}}.
	
	
	
	\bibitem{aranda2012} J. I. Aranda, J. Monta\~no, F. Ram\'irez-Zavaleta, J. J. Toscano, and E. S. Tututi, \href{https://doi.org/10.1103/PhysRevD.86.035008
	}{\textcolor{blue}{Phys. Rev. D \textbf{86}, 035008 (2012)}}.
	
	
	\bibitem{FVcms} The CMS Collaboration, \href{https://doi.org/10.48550/arXiv.2205.06709
	}{\textcolor{blue}{arXiv:2205.06709}};  \href{https://doi.org/10.1007/JHEP04(2018)073
	}{\textcolor{blue}{JHEP \textbf{04}, 073 (2018)}}.
	
	
	\bibitem{FVatlas} M. Aaboud \textit{et al}. (ATLAS Collaboration), \href{https://doi.org/10.1103/PhysRevD.98.092008
	}{\textcolor{blue}{Phys. Rev. D \textbf{98}, 092008 (2018)}}.
	
	
\bibitem{PX} Hiren H. Patel, \href{https://doi.org/10.1016/j.cpc.2015.08.017}{\textcolor{blue}{Comput. Phys. Commun. \textbf{197}, 276 (2015)}}.
	
	
\bibitem{Gi98} G.-C Cho, K. Hagiwara and Y. Umeda,
\href{https://doi.org/10.1016/S0550-3213(98)00549-5}{\textcolor{blue}{Nucl. Phys. \textbf{B531}, 65 (1998)}}.
	
\bibitem{PJW} P. Langacker and J. Wang,
\href{https://doi.org/10.1103/PhysRevD.58.115010}{\textcolor{blue}{Phys. Rev. D \textbf{58}, 115010 (1998)}}.



\bibitem{durkin} L. S. Durkin and P. Langacker, \href{https://doi.org/10.1016/0370-2693(86)91594-7}{\textcolor{blue}{Phys. Lett. \textbf{B166}, 436 (1986)}};
M. Cvetic and P. Langacker, Proceedings of Ottawa 1992:  Beyond the standard model 3, 454-458, (1992); C-W Chiang, Yi-Fan Lin, and Jusak Tandean, \href{https://doi.org/10.1007/JHEP11(2011)083}{\textcolor{blue}{JHEP \textbf{11}, 083 (2011)}}.
	
\bibitem{RW82} R. W. Robinett, J. L. Rosner,
\href{https://doi.org/10.1103/PhysRevD.25.3036}{\textcolor{blue}{Phys. Rev. D \textbf{25}, 3026 (1982)}}; Erratum
\href{https://doi.org/10.1103/PhysRevD.27.679}{\textcolor{blue}{Phys. Rev. D \textbf{27}, 679 (1983)}}.


\bibitem{langacker3} P. Langacker and M. Pl\"umacher, \href{https://doi.org/10.1103/PhysRevD.62.013006}{\textcolor{blue}{Phys. Rev. D \textbf{62}, 013006 (2000)}};
X.-G. He and G. Valencia, \href{https://doi.org/10.1103/PhysRevD.74.013011}{\textcolor{blue}{Phys. Rev. D \textbf{74}, 013011 (2006)}};
C.-W. Chiang, N. G. Deshpande, and J. Jiang, \href{https://doi.org/10.1088/1126-6708/2006/08/075}{\textcolor{blue}{JHEP \textbf{08}, 075 (2006)}}.


\bibitem{mc} M. Cveti\text{$\check{c}$}, B. Kayser and P. Langacker,
\href{https://doi.org/10.1103/PhysRevLett.68.2871}{\textcolor{blue}{Phys. Rev. Lett. \textbf{68}, 2871 (1992)}}.


\bibitem{Fa} F. del Aguila, Acta Phys. Pol. B \textbf{25}, 1317 (1994);
\href{https://doi.org/10.48550/arXiv.hep-ph/9404323}{\textcolor{blue}{arXiv: hep-ph/9404323}}.	
		
\bibitem{Salam-Mohapatra} J. C. Pati and A. Salam, \href{https://doi.org/10.1103/PhysRevD.10.275}{\textcolor{blue}{Phys. Rev. D \textbf{10}, 275 (1974)}}; Erratum \href{https://doi.org/10.1103/PhysRevD.11.703.2}{\textcolor{blue}{Phys. Rev. D 11, 703 (1975)}};
	
\bibitem{david} J. I. Aranda, E. Cruz-Albaro, D. Espinosa-G\'omez, J. Monta\~no, F. Ram\'irez-Zavaleta, and E. S. Tututi,
\href{https://doi.org/10.1142/S0217751X21501670}{\textcolor{blue}{Int. J. Mod. Phys. A\textbf{36}, 2150167 (2021)}}.
	


	
	
	\bibitem{kuno} Y. Kuno and Y. Okada, \href{https://doi.org/10.1103/RevModPhys.73.151}{\textcolor{blue}{Rev. Mod. Phys. \textbf{73}, 151 (2001)}}.
	
	\bibitem{af} A. Faessler, T. Gutsche, S. Kovalenko, V. E. Lyubovitskij, and I. Schmidt, \href{https://doi.org/10.1103/PhysRevD.72.075006}{\textcolor{blue}{Phys. Rev. D \textbf{72}, 075006 (2005)}}.
	
	\bibitem{kosmas} T. S. Kosmas, G. K. Leontaris,  and J. D. Vergados, \href{https://doi.org/10.1016/0146-6410(94)90047-7}{\textcolor{blue}{Prog. Part. Nucl. Phys.  \textbf{33}, 397 (1994)}}.
	
	\bibitem{Jbernabeu} J. Bernab\'eu, E. Nardi and D. Tommasini, \href{https://doi.org/10.1016/0550-3213(93)90446-V}{\textcolor{blue}{Nucl. Phys. \textbf{B 409}, 69 (1993)}}.
	
	\bibitem{HC2} H. C. Chiang, E. Oset, T. S. Kosmas, A. Faessler and J. D. Vergados, \href{https://doi.org/10.1016/0375-9474(93)90259-Z}{\textcolor{blue}{Nucl. Phys. A \textbf{559}, 526 (1993)}}.
	
	
	
	\bibitem{brenda}  J. Monta\~no-Dom\'inguez, B. Quezadas-Vivian, F. Ram\'irez-Zavaleta, and E.S. Tututi, \href{https://doi.org/10.1088/1361-6471/ac69ff}{\textcolor{blue}{J. Phys. G \textbf{49}, 075004 (2022)}}.
	
	\bibitem{Suzuki} T. Suzuki, D.F. Measday, and J.P. Roalsvig, \href{https://doi.org/10.1103/PhysRevC.35.2212}{\textcolor{blue}{Phys. Rev. C \textbf{35}, 2212 (1987)}}.
	
	
	\bibitem{HC} H. C. Chiang, E. Oset and P. Fern\'andez de C\'ordoba, \href{https://doi.org/10.1016/0375-9474(90)90350-U}{\textcolor{blue}{Nucl. Phys. A \textbf{510}, 591 (1990)}}.
	
	\bibitem{sundrumII} C. Dohmen, et al, SUNDRUM II Collab. \href{https://doi.org/10.1016/0370-2693(93)91383-X}{\textcolor{blue}{ Phys. Lett. \textbf{B 317}, 631 (1993)}}.
	
	\bibitem{Faguila} F. del Aguila, J. de Blas, and M. Perez-Victoria, \href{ 	
		https://doi.org/10.1007/JHEP09(2010)033}{\textcolor{blue}{JHEP 09, 033 (2010)}}, \href{   	
		https://doi.org/10.48550/arXiv.1005.3998}{\textcolor{blue}{ 	arXiv:1005.3998}}.
	
	
	
	
	
	\bibitem{aranda2011} J. I. Aranda, F. Ram\'irez-Zavaleta, J. J. Toscano, and E. S. Tututi, \href{https://doi.org/10.1088/0954-3899/38/4/045006}{\textcolor{blue}{J. Phys. G: Nucl. Part. Phys. \textbf{38}, 045006 (2011)}}.
	
	
\end{thebibliography}
\end{document}